\documentclass[twocolumn,showpacs,showkeys,preprintnumbers,amsmath,amssymb]{revtex4}
\pdfoutput=1
\usepackage{graphicx}% Include figure files
\usepackage{dcolumn}% Align table columns on decimal point
\usepackage{bm}% bold math

\usepackage{color}
\usepackage{xcolor}

\begin{document}
 
%\preprint{APS/123-QED}

\title{Interaction of ultrashort X-ray pulses with \boldmath$\mathrm{B_{4}C}$, SiC and Si}% Force line breaks with \\

\author{M. Bergh}
\email{magnus@xray.bmc.uu.se}
\affiliation{Laboratory of Molecular Biophysics, Uppsala University,\\Husargatan 3, Box 596, SE-75124, Uppsala, Sweden}
%\altaffiliation[Also at ]{Laboratory of Molecular Biophysics, Uppsala University,\\Husargatan 3, Box 596, SE-75124, Uppsala, Sweden}%Lines break automatically or can be forced with \\

\author{S. P. Hau-Riege}
\affiliation{Lawrence Livermore National Laboratory, Livermore, California 94551, USA}

\author{H. A. Scott}
\affiliation{Lawrence Livermore National Laboratory, Livermore, California 94551, USA}

\author{N. T\^\i mneanu}
\email{nicusor@xray.bmc.uu.se}
\affiliation{Laboratory of Molecular Biophysics, Uppsala University,\\Husargatan 3, Box 596, SE-75124, Uppsala, Sweden}

%\affiliation{

\date{\today}% It is always \today, today,

\begin{abstract}
The interaction of 32.5 and 6 nm ultrashort X-ray pulses with the solid materials 
$\mathrm{B_{4}C}$, SiC and Si is simulated with a non-local thermodynamic equilibrium (NLTE) 
radiation transfer code. We study the ionization dynamics as function of depth in the material, 
modifications of the opacity during irradiation and estimate crater depth. Furthermore, 
we compare the estimated crater depth with experimental data, for fluences up to 2.2 
$\mathrm{J/cm^2}$. Our results show that at 32.5 nm irradiation, the opacity changes with 
less than a factor of 2 for $\mathrm{B_{4}C}$ and Si and a factor of 3 for SiC, 
for fluences up to 200 $\mathrm{J/cm^2}$. At a laser wavelength of 6 nm, the model 
predicts a dramatic decrease in opacity due to the weak inverse bremsstrahlung, 
increasing the crater depth for high fluences.  
\end{abstract}

                            
\maketitle

\section{introduction}

Novel laser sources that can generate extremely intense ultrashort X-ray pulses are developing fast. The Free Electron Laser in Hamburg (FLASH) \cite{Ackermann2007} has produced soft X-ray pulses at 1e14 $\mathrm{W/cm^2}$ with a 20 micrometer focus at a wavelength of 32.5 nm, and will reach 6 nm by the end of 2007 \cite{Ayvazyan2006}. At SLAC in California, the Linac Coherent Light Source (LCLS) \cite{LCLS} is expected to deliver ultrashort pulses of hard X-rays in 2008. 

Methods developed at large-scale facilities are likely to become more accessible due to the advance of table top sources with similar characteristics; high harmonic generation \cite{McPherson1987,Ferray1988} has demonstrated 25 fs pulses at $\mathrm{10^{14}W/cm^2}$ \cite{Takahashi2000}, and laser wake-field accelerators producing mono-energetic electron beams \cite{Leemans2004,Faure2004,Gruner2007} may act as a source for ultrashort X-rays.

These lasers will access an unexplored regime for light-matter interaction, extending the capabilities of existing scientific techniques and enabling the development of new ones. Laser ablation is of great interest partly as a tool for processing material with nanometer precision, but also as a way of testing microscopic theories describing the interaction of X-ray laser pulses with solids. Predictive models could anticipate new applications and illuminate the limitations set by radiation damage in a sample and in focusing optics. Coherent diffraction imaging of biological specimens using free-electron lasers \cite{Neutze2000a,Bergh2007} is an example of an application where the achievable resolution depends on the ultrafast ionization dynamics of the X-ray-solid interaction. 

The interaction of ultrashort optical laser pulses with solids has been studied extensively experimentally \cite{Tinten1995,Lindenberg2005a} and theoretically \cite{Town1994,Jiang2004,lorazo2006}. However, for VUV/X-ray pulses the dominant physical processes that govern the interaction are different, resulting in a quite different response of the material. \citet{Fajardo2004} discuss the mechanisms of laser absorption for photon energies in the range of tens to a hundred eV, and pulse lengths of 1 ps to 1 ns.  They present simulations of hydrodynamic expansion for different pulse lengths and target materials. \citet{Theobald1999} has investigated the optical properties of dense plasmas in the extreme ultraviolet spectral range using high harmonic generation.  

Recently, experiments at FLASH have been performed using a holographic time-delay technique \cite{Chapman2007b}. The setup realized unique femtosecond time-resolved measurements of the structural changes induced by the FEL-material interaction. The modeling tools used to reproduce the experimental results from \cite{Chapman2007b} are presented in more depth by Hau-Riege et al. \cite{hau-riege2007c}, where the optical properties of dense plasmas are investigated assuming local thermodynamic equilibrium (LTE) conditions.  

In this article we take a step further and present non-local thermodynamic equilibrium (NLTE) simulations of the heating of solids by femtosecond VUV and X-ray lasers at different fluences. Three different materials are considered: $\mathrm{B_{4}C}$, SiC and Si. These materials are interesting to compare due to their differences in Z and electronic structure, which result in quite different responses to intense X-ray pulses. The materials are also candidates for X-ray optics, and the ablative properties of semiconductors are crucial in micro-processing applications. First, we present simulations of VUV laser heating experiments \cite{Hau-Riege2007d} that were performed at FLASH. We define a threshold for ablation to identify ablation regions in the simulations and compare to experimental crater depths. We then investigate how absorption mechanisms, and thereby depth profiles of heated targets, would change for the higher fluences which could be achieved at the FLASH-facility under tight focusing conditions. Finally, we present simulations for pulses around 6 nm wavelength to investigate the ionization dynamics for higher photon energies.

\section{model description}

The interaction of intense ultrafast lasers with solid material is a complicated process. The plasma resulting from such an interaction is a transient state of matter that is exceptionally hard to probe. Ultrafast X-ray sources grant access to this regime by combining a short wavelength, which can propagate through the dense plasma, with a pulse length short enough to probe its temporal evolution \cite{Theobald1996,Dobosz2005}. Ultrafast X-ray sources are thus important both as primary pump pulses in applications dealing with imaging \cite{chapman2006a} and material processing \cite{Chalupsky2007}, and as secondary pulses that can probe the region of interaction. A theoretical framework can either start from the solid state formulation (see e.g. \cite{Stampfli1994}) or from a plasma formulation originally developed for low-density hot matter \cite{Eidmann2000,Fajardo2004}. The former approach is suitable for fluences around the melting threshold, while the latter is suitable for higher fluence where ionization is considerable and chemical bonds have little significance. For the present study we choose a plasma formulation with high-density corrections.   

The laser-matter interaction is modeled in Cretin \cite{scott1992,scott2001}, a multi-dimensional NLTE radiation transfer code. The calculated level populations and transition rates provide opacities, heating rates and conduction coefficients in each time step of the simulation. For this study we use the code's built-in screened hydrogenic atomic models, which are similar to those described in \citet{More1982}. The cold opacity  has been scaled to fit the Henke absorption coefficients \cite{Henke1993} at the laser wavelength for the specific materials. The electron energy distribution is assumed to be thermal, i.e. the electrons thermalize instantaneously, so transition rates dependent on the electron energy distribution are obtained by integrating over a Maxwellian electron distribution. A study using the Fokker-Planck formalism \citep{Town1994} for laser heated solids at similar temperatures and time scale concludes that the effect of a non-Maxwellian distribution is small.

To account for the effects of high density, ionization potentials are lowered following the Stewart-Pyatt formula \citep{stewart1966}. This is a commonly used approximate model and the results have been compared to those of more detailed models and to experimental emission spectra \citep{nantel1998}. For solid densities, this continuum lowering has a considerable effect on the ionization dynamics. In the case of a 32.5 nm, 25 fs laser pulse on SiC, the average ionization (zbar) is almost doubled at an intensity of 1e15 W/cm2.

The inverse bremsstrahlung process (free-free absorption) applies the formalism of \citet{Dawson1962}. The Coulomb logarithm uses a maximum impact parameter $b_{max}= \sqrt{{\lambda_D}^2+{a_i}^2}$ suggested by \cite{Schlanges2002} for dense systems. ${\lambda_D}$ is the Debye length and $a_i=(3/4\pi n_i)^{1/3}$ is the ion sphere radius and $n_i$ is the ion density. The electron-ion coupling coefficient is calculated from Spitzer's formula \cite{Spitzer1956}, using the same Coulomb logarithm as above.  

We model the heating in one dimension, justified by the laser spot size being much larger than the penetration depth into the material. The simulations use 40 zones which exchange energy through radiation transport and electron thermal conduction. The incident radiation is attenuated as it passes through each zone, and the opacity of each zone is calculated each time-step from the atomic populations and the inverse bremsstrahlung model. 

In the case of optical laser heating, a critical electron density for the propagation of the laser light is usually reached in an early phase of the the exposure and the laser energy is deposited mainly in a surface layer. The main channels of ionization are multi-photon ionization (MPI) and avalanche breakdown, where delocalized electrons gain energy through inverse bremsstrahlung followed by a geometric increase of free electrons through impact ionization. For intensities above $\mathrm{\approx 10^{15} W/cm^2}$ the optical laser field becomes strong enough to suppress the atomic/ionic potential, causing field ionization. The Keldysh parameter (for a review, see e.g. \cite{Brabec2000}) indicates which process is dominant. For the short wavelengths considered in this study the Keldysh parameter $\gamma>1$, indicating that field ionization can be neglected. To estimate the degree of MPI, the simulations include below-threshold ionization with weak-field scalings from \citet{Delone2000}.

%***********************************
%       RESULTS AND DISCUSSION
%***********************************

\section{results and discussion}

In this section we present comparison to experimental data for the materials $\mathrm{B_{4}C}$, SiC and Si, as well as predictions of the response of the materials at higher fluence and photon energies than the current FEL beam parameters. The theory presented in the modeling section is used to simulate the ionization dynamics throughout the material. The laser beam parameters have been chosen to match measured \cite{Ackermann2007} and predicted \cite{Salin2006} values at the FLASH facility. 

\begin{figure}[htb]
\includegraphics[width=7cm]{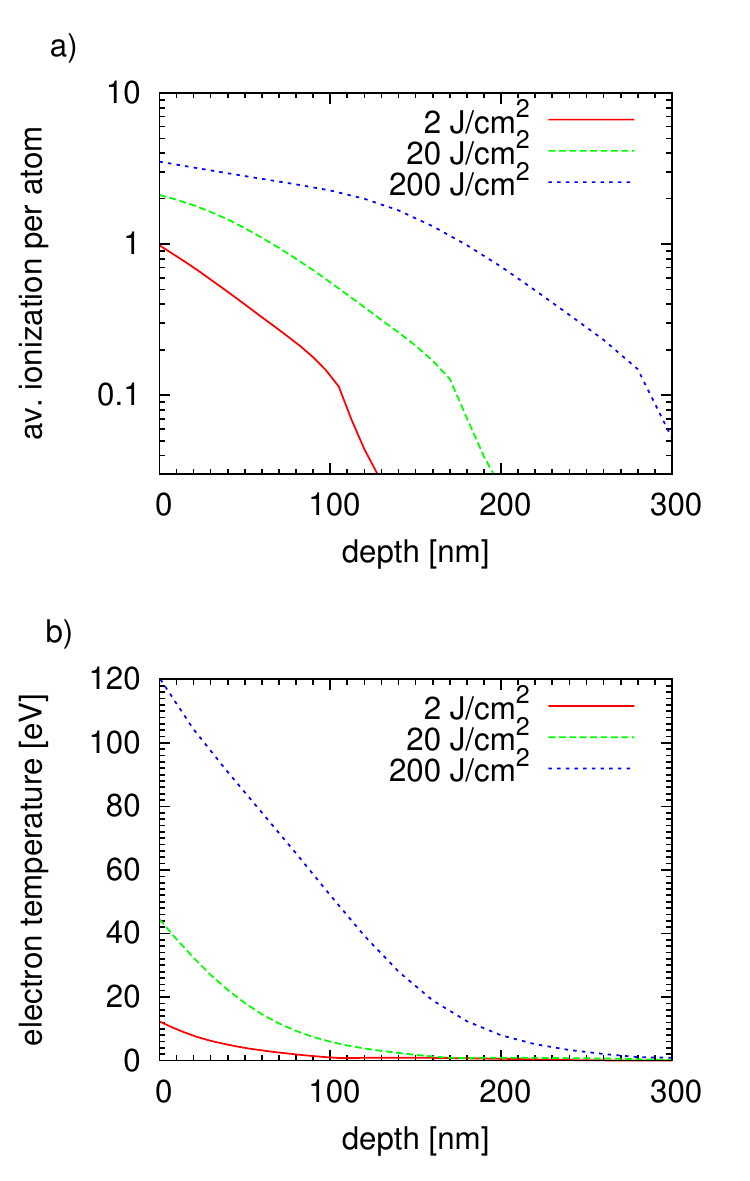}
\caption{The average ionization per atom (a) and electron temperature (b) after the laser pulse as a function sample depth for B4C at different fluences.}\label{fig:zbar}
\end{figure}

\subsection{Crater depth and mechanisms of ablation}

Figure \ref{fig:zbar} shows the simulated average ionization (a) and electron temperature (b) as functions of depth for  $\mathrm{B_{4}C}$ at different fluences. The temporal structure of the pulse is chosen to match the pulse parameters at FLASH and is  modeled by a Gaussian distribution with a FWHM of 25 fs, a wavelength of 32.5 nm, and with a spectral width of 1\%. The intensity achieved so far in experiments at FLASH is similar to the red curves in figure \ref{fig:zbar}, but we note that under tight focusing conditions, fluences in parity with the blue curve can be expected. The damage mechanism for VUV and X-ray irradiation in solid materials proceeds through different processes compared to that for optical lasers. In the latter case, valence electrons are excited to the conduction band, and as these free carriers accumulate they can be described as an electron gas \cite{Stampfli1994}. Lorentzo et al. \cite{lorazo2006} find that the optical laser-excited carriers at near-threshold ablation in silicon reach an equilibrium at around 0.8 eV through mutual collisions and impact ionization. On the other hand, for photon energies well above the binding energy of the valence electrons, photoelectrons are ejected into the lattice, generating many secondary electrons through impact ionization. This is a fast process, where an X-ray photon has generated a cascade of secondary electrons in a few femtoseconds \cite{Ziaja2005a,Timneanu2004a}. At a depth in the material where the beam is attenuated to a fluence approaching the threshold for ablation, the simulations suggest that the state of the material is similar to the optical case; the quasi-free electrons here, produced mainly through impact ionization, reach a temperature of 0.5-1.5 eV. The corresponding average ionization is around 0.1 electron per atom. The simulations also show that the electron-ion equilibration time is much longer than the pulse, approximately 1 picosecond. Melting and ablation that occurs on a time-scale faster than conventional melting is usually referred to as non-thermal melting \cite{Stampfli1994}. Calculations \cite{Stampfli1994} and measurements \cite{Rousse2001} for Si and GaAs indicate that the threshold can be expressed as the density of free carriers required to destabilize the lattice, and that its value is $N_{uf}\approx\mathrm{10^{22} cm^{-3}}$. In the simulations in this study, the depth where the density of free electrons reaches this threshold value corresponds to an electron temperature that lies within the range of critical temperatures of solids of 0.3-0.9 eV \cite{More1988,Martynyuk1974}. This motivates us to identify this region of electron density and temperature as a threshold for material removal. Thus, the simulations in figure \ref{fig:zbar} yield the depth into the material where we predict non-thermal ablation will occur. We note that recent studies of damage from the FLASH free-electron laser \cite{Stojanovic2006,Chalupsky2007} present experimental data supporting the idea that material removal proceeds via non-thermal ablation.

%%%%%%%%%%%%%%%%%%%%%%%%%%%%

%DEPTH COMPARISON

%%%%%%%%%%%%%%%%%%%%%%%%%%%%

\begin{figure}[htb]
\includegraphics[width=7cm]{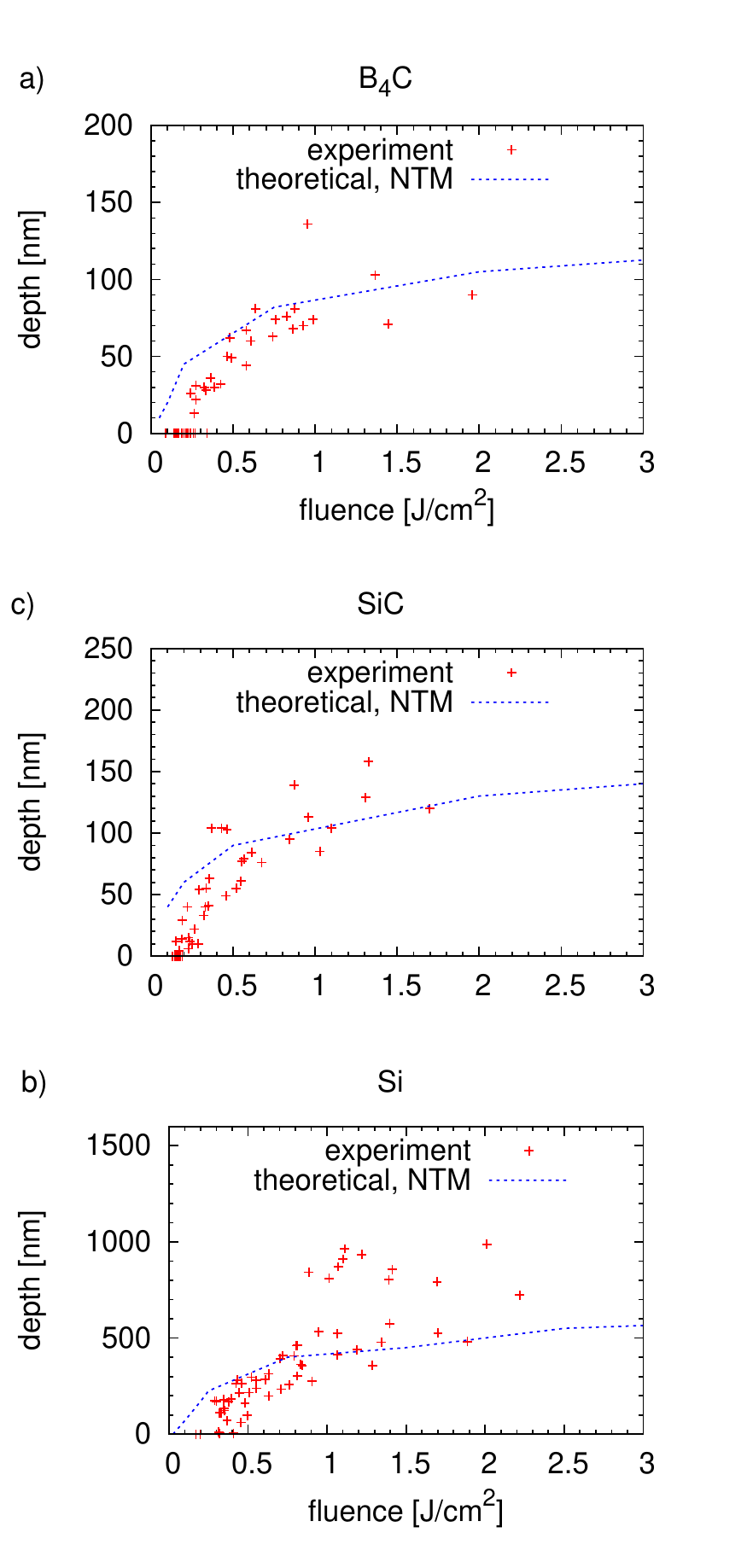}
\caption{Experimental crater depth as function of fluence (red) compared to heating simulations using the threshold for ultrafast melting (UFM) (blue), for $\mathrm{B_{4}C}$ (a), SiC (b) and Si (c). For $\mathrm{B_{4}C}$ and SiC support the notion that material removal is initiated by ultrafast melting. In the case of Si, expulsion of melted material could be significant for fluences above 1 $\mathrm{J/cm^2}$.}\label{fig:depth}
\end{figure}

 Figure \ref{fig:depth} shows a comparison between simulated and measured crater depths, with measured depths obtained from post mortem analysis of X-ray free-electron laser irradiated samples. The conditions of the experiment are described in more detail by \citet{Hau-Riege2007d}. The blue curves show that the predicted depths for $\mathrm{B_{4}C}$, SiC and Si. For $\mathrm{B_{4}C}$ and SiC the agreement is good, even though the data is scarce above 1 $\mathrm{J/cm^2}$. For fluences close to the ablation threshold, the simulated depth is larger than the measured values for all materials. We note that the plasma formulation does not reproduce the thresholds for damage and ablation accurately; that would require a solid-state approach. In this study the aim is to quantify the response of the material during irradiation, and to estimate the crater depths for higher fluences. For Si the agreement is decent up to a fluence of 1 $\mathrm{J/cm^2}$, above which the experimental crater depths vary quite a lot. A similar effect was recently reported by \citet{Stojanovic2006}, who observed clean craters in Si at 0.9 $\mathrm{J/cm^2}$, but a drastic increase in depth and a more complicated crater structure at 1.5 $\mathrm{J/cm^2}$. These findings, together with a time resolved study of optical reflectivity changes during heating and ablation, support the notion that the ejection of molten material becomes an important mechanism in Si above a certain fluence. They also point out that this "piston effect" is expected to be more pronounced for large absorption depths, which suggest that the such an effect would be less pronounced in $\mathrm{B_{4}C}$ compared to Si. The crater depth data in figure \ref{fig:depth}(b) show a sudden increase and spread in depth above 1 $\mathrm{J/cm^2}$. In line with the observations above, we assume that this effect is due to the expulsion of melted material. Hence, Si is excluded from the crater depth predictions at higher fluences. 

%An accurate model for calculating the threshold for melt expulsion lies outside the scope of this article; it would require a reliable equation of state model in the warm dense matter region and modeling of hydrodynamic instabilities.  

\begin{figure}[htb]
\includegraphics[width=7cm]{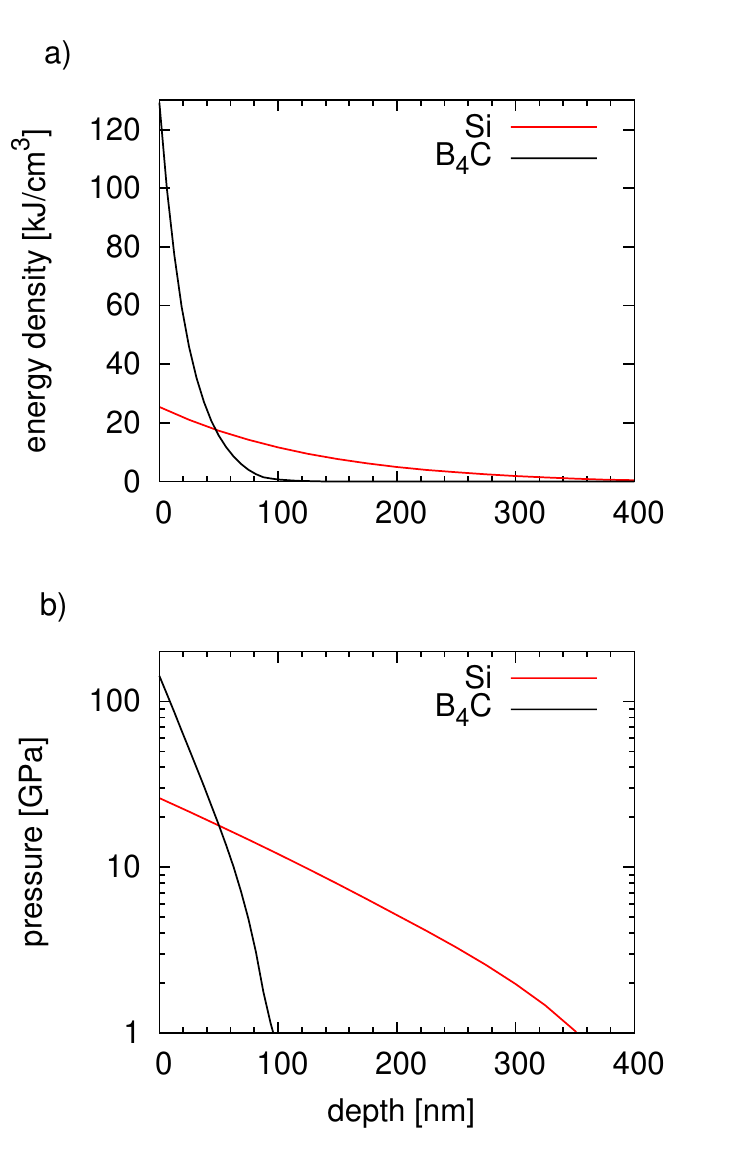}
\caption{(a) The deposited energy density as function of depth for $\mathrm{B_{4}C}$ (black curve) and Si (red curve) at the end of a 1 $\mathrm{J/cm^2}$, 25 fs, 32.5 nm pulse. An exponential decay with depth indicates that the opacity of the material remains close to constant during the pulse. (b) Total electron pressure in the materials as function of depth.}\label{fig:si_press}
\end{figure}

Figure \ref{fig:si_press} shows two quantities that are important for the ablation mechanisms in these materials. In (a) the deposited energy density at the end of a 1 $\mathrm{J/cm^2}$ pulse  is plotted as a function of depth. The steep profile for $\mathrm{B_{4}C}$ (black curve) compared to Si (red curve) reflects the larger penetration depth of Si. The higher energy density at the surface in $\mathrm{B_{4}C}$ is a result of the higher ionization and electron temperature. This leads to a hydrodynamic pressure that reaches 120 GPa, which is almost 6 times higher than that for Si (b). For both materials, the pressure at the end of the pulse is dominated by the electrons, with the ion and radiation pressure being less than 3\% of the total pressure. The pressure at the $\mathrm{B_{4}C}$ surface is higher, but for Si there is more material with a temperature around the melting/boiling point due to the flat heating profile. In combination with the lower heat capacity of Si, this could possibly enhance the effect of melt expulsion.

In the case of $\mathrm{B_{4}C}$ and SiC, the amount of melt expulsion appears to be considerably lower than for the Si case, and it has been suggested that the removal of material happens through two-phase vaporization followed by fluid-like expansion \cite{Hau-Riege2007d}. Large-scale hydrodynamic motion of 140 nm spherical polystyrene particles under the same beam conditions as in figure \ref{fig:depth} has been confirmed to take place during the first picoseconds \cite{Chapman2007b}. During this time, our simulations indicate that further heating through electron thermal conduction and radiation transport from the hot region is negligible. However, such a sudden acceleration of material from the surface will release a compressive shock wave into the material \cite{Eidmann2000,Fajardo2004,hau-riege2007c}. To estimate the degree of heating in the material after the pulse, we simulate the material beneath the vaporized expanding material for 10 ps, and apply a Gaussian compression of the density (with a factor of 2 as reported for similar conditions \cite{Fajardo2004}) over time as a crude model of a shock wave. The result suggests that there is very little further ionization after the pulse for the fluences considered in this study, and that the calculated ionization depth-profile after the pulse can be used to estimate the crater depth when vaporization followed by fluid-like expansion is the dominant mechanism.

%The electron-ion coupling is followed using the Spitzer HLS model \cite{Schlanges2002}. The coupling constant is also calculated using an expression used by \cite{Eidmann2000}, derived for degenerate matter below the Fermi temperature. We also follow recombination, electron thermal conduction and radiation transport. 

%%%%%%%%%%%%%%%%%%%%%%%

%OPACITY INVESTIGATION

%%%%%%%%%%%%%%%%%%%%%%%
 
\subsection{Predictions at high fluence}

From the analysis above, we see that the plasma model shows reasonable agreement with the experimental crater depths for $\mathrm{B_{4}C}$ and SiC for fluences of 0.5-2 $\mathrm{J/cm^2}$. Si is included in the study of the material response \textit{during} the pulse, but excluded in the prediction of the crater depth at high fluences due to the complicated ablative properties. At higher fluences, the optical properties in the X-ray regime can be expected to change due to significant changes in the electronic structure. Such modifications can be studied by following the atomic populations in the constituent atoms. Below, we present predictions for the behavior of these materials during irradiation from pulses with fluences up to 500 $\mathrm{J/cm^2}$. This corresponds to the order of magnitude that can be expected at the FLASH facility under tight focusing conditions, and at future X-ray sources.  

%Simulations of the temporal structure of the pulse \cite{Salin2006}, predict that the pulse will contain spikes that results in instantaneous intensities higher than a Gaussian pulse with the same total energy and pulse length. To investigate the effect of such intensity variations on the ionization dynamics, we did test simulations and monitored the electron temperature on the surface of the materials for both Gaussian and spiky pulse shapes, where the temporal structure of the spiky pulse is made to resemble the predicted structures in  \cite{Salin2006}. The result suggests that for 32.5 nm wavelength and at fluences below 500 $\mathrm{J/cm^2}$, the temperatures after the pulse differ by less than 1\%. This procedure is described in more depth in section C. 

\begin{figure}[htb]
\includegraphics[width=7cm]{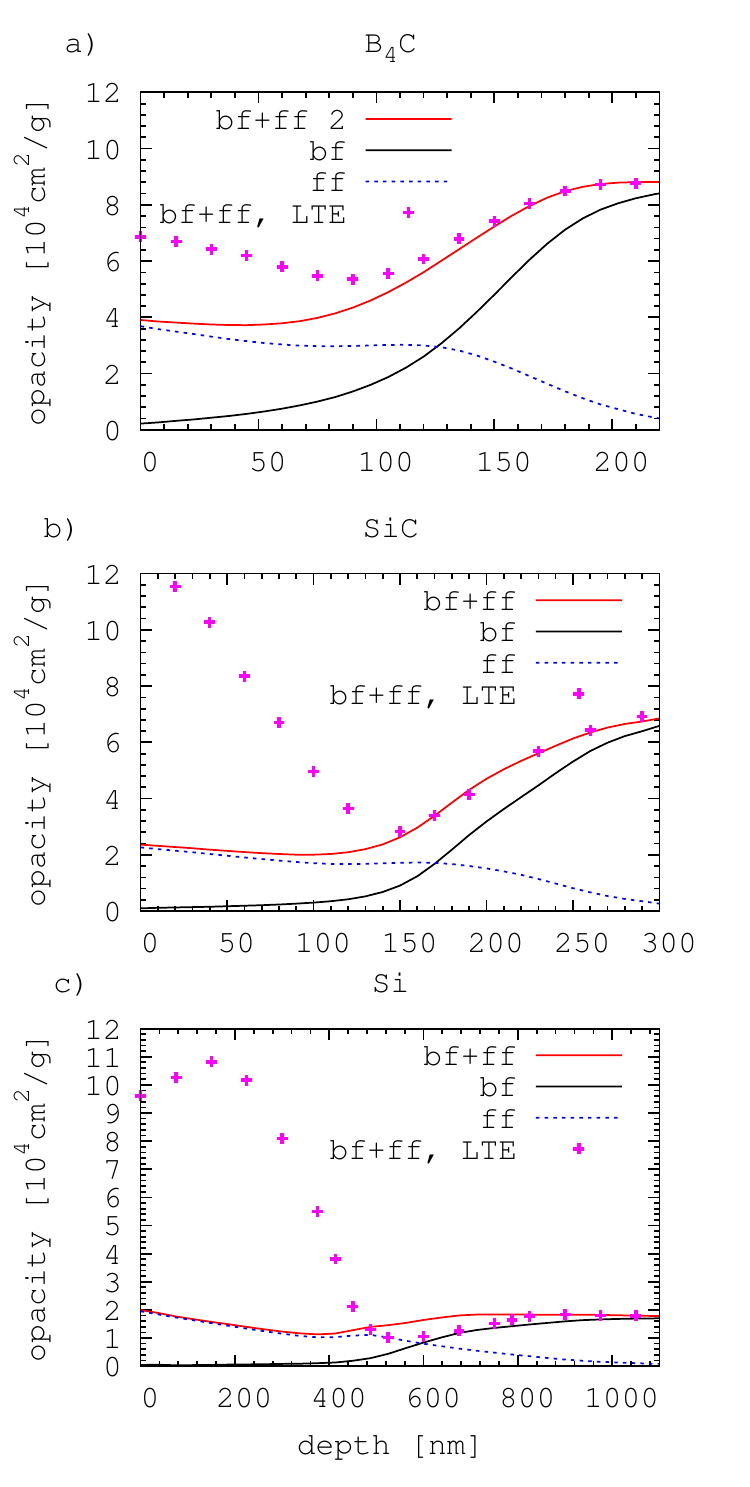}
\caption{Opacity as function of sample depth at the end of a 25 fs, 32.5 nm pulse. The fluence is 200 $\mathrm{J/cm^2}$ in the case of $\mathrm{B_{4}C}$ (a) and SiC (b), and 500 $\mathrm{J/cm^2}$ in the case of Si (c). The non-LTE total opacity (bf+ff, red curve) is the sum of the free-free (ff) contribution (blue) and the bound-free contribution (black). The bf opacity curve converges to the cold opacity at large depths, whereas the ff opacity is sensitive to the number of free electrons and hence increase closer to the surface.  For comparison, the dotted curve shows the total opacity assuming LTE, at the same temperatures as in the non-LTE case. The LTE opacity model results in a more restrained ionization at low temperatures, whereas at high temperatures the LTE ionization is stronger, increasing the inverse bremsstrahlung absorption.}\label{fig:op_lte}
\end{figure}

The absorption of the laser light in the material is modified due to plasma formation during the pulse. This is illustrated in figure \ref{fig:op_lte}, which shows the opacity at the laser wavelength of 32.5 nm as a function of depth at the end of an 200 $\mathrm{J/cm^2}$ pulse for $\mathrm{B_{4}C}$ (a) and SiC (b), and for a 500 $\mathrm{J/cm^2}$ pulse for  Si (c). These fluences heat the materials to comparable electron temperatures at the end of the pulse ($\mathrm{\approx}$ 125 eV). At a laser wavelength of 32.5 nm there is no contribution from the bound-bound transitions for these elements, so the opacity comes from bound-free (bf) and free-free (ff) transitions. The right hand side of the plots corresponds to a depth where the material is cold and the opacity is proportional to the initial photoionization cross section.  As we move left toward the surface, the bf-contribution decreases as the atoms ionize, and the ff-contribution increases as a result of the increase in the density of quasi-free electrons. In the case of $\mathrm{B_{4}C}$, only the core electrons remain in the zones closest to the surface, leading to a flattening of the opacity curve (bf+ff) with depth. The depth profile is similar for SiC. In the case of pure Si, the opacity starts to rise close to the surface. This is because the L-shell electrons in Si have a lower ionization potential than the K-shell electrons in C and B, producing a higher density of quasi-free electrons for the same temperature, and hence a stronger ff-contribution. It is of interest to find the fluence where the opacity starts to deviate from the cold value. For $\mathrm{B_{4}C}$ and SiC the opacity changes by 25\% at about 5 $\mathrm{J/cm^2}$. For Si the fluence is about 20 $\mathrm{J/cm^2}$. We also note that the free-free contribution becomes larger than the bound-free contribution at electron temperatures of 35, 31 and 23 eV for $\mathrm{B_{4}C}$, SiC and Si respectively. 

To emphasize the difference in ionization dynamics compared to an LTE treatment, figure  \ref{fig:op_lte} also shows the opacity resulting from a Saha-Boltzmann population distribution at the same temperatures as in the non-LTE case. The main reason for the differences is that the LTE treatment results in a higher average ionization, leading to enhanced inverse bremsstrahlung absorption. The overestimation of the average ionization in LTE is known from previous studies \cite{Mima1994,Butler2004}. By performing the same analysis at lower fluences we can find where the departure from LTE begins to have an impact on the opacity at the surface. We choose to consider a difference in opacity of 25\% as "possible to distinguish" in, e.g. a transmission measurement. This is reached at an electron temperature of 45 eV and 92 eV for Si and $\mathrm{B_{4}C}$ respectively, validating the LTE treatment used in \cite{hau-riege2007c}. These temperatures are reached for similar fluences; 65 and 90 $\mathrm{J/cm^2}$ respectively. We note that this result holds for similar pulse lengths and wavelengths only; an estimate of the deviation from LTE must be found whenever the pulse parameters change significantly. 

\begin{figure}[htb]
\includegraphics[width=7cm]{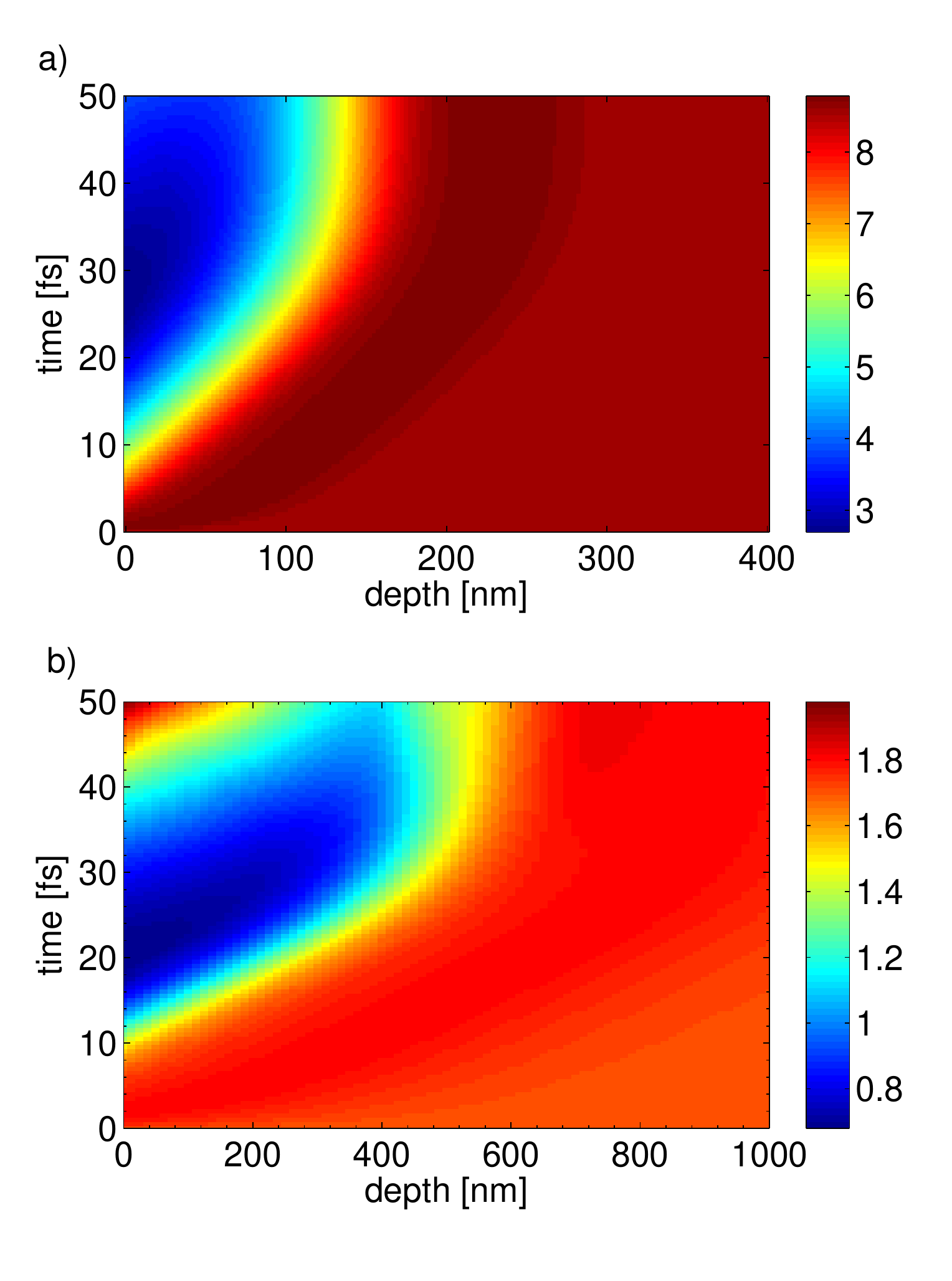}
\caption{The opacity (in $\mathrm{cm^2/g}$) for $\mathrm{B_{4}C}$ (a) and Si (b) as a function of material depth and time during pulse irradiation, for the same pulse parameters as in figure \ref{fig:op_lte}. The materials reach a similar electron temperature at the surface at the end of the pulse. The opacity is rapidly reduced by about a factor of two in the surface region for both materials, but in the case of Si it reaches the initial value at the end of the pulse due to the higher rate of inverse bremsstrahlung heating.}\label{fig:op_b4c_si}
\end{figure}

%for $\mathrm{B_{4}C}$ (a) and Si (b), with fluences of 200 $\mathrm{J/cm^2}$ and  500 $\mathrm{J/cm^2}$ respectively.

The heating depth profile after the pulse depends on the opacity throughout the interaction. Figure \ref{fig:op_b4c_si} illustrates the temporal evolution of the opacity during the pulse as function of depth for $\mathrm{B_{4}C}$ (a) and Si (b), with the same fluences as in figure \ref{fig:op_lte}. The SiC depth-time profile is similar to that of $\mathrm{B_{4}C}$.  The material in the blue-colored regions attenuates less radiation, but due to the inverse bremsstrahlung contribution, the opacity doesn't decrease by more than a factor of two. In the case of Si, the density of free electrons is higher, even though the temperature in the region close to the surface is similar. This leads to increased inverse bremsstrahlung absorption, increasing the opacity at the surface during the last 10 fs of the simulation. Figure \ref{fig:op_b4c_si} also shows the depth at which a significant modification of the opacity can be expected. We note that by choosing a sample with a thickness similar to this depth (e.g. a thin film), one could measure the transmission of either the pump-beam or of a probe beam arriving at a later time. Such a set-up is straightforward in principle, and could be useful for testing NLTE predictions, or for validating existing models for inverse bremsstrahlung in solid density plasmas. Furthermore, by probing the region of interaction after heating it with a second pulse, one could measure the electron density. Such a measurement could contribute to the process of validating hot dense plasma models.     

\begin{figure}[htb]
\includegraphics[width=7cm]{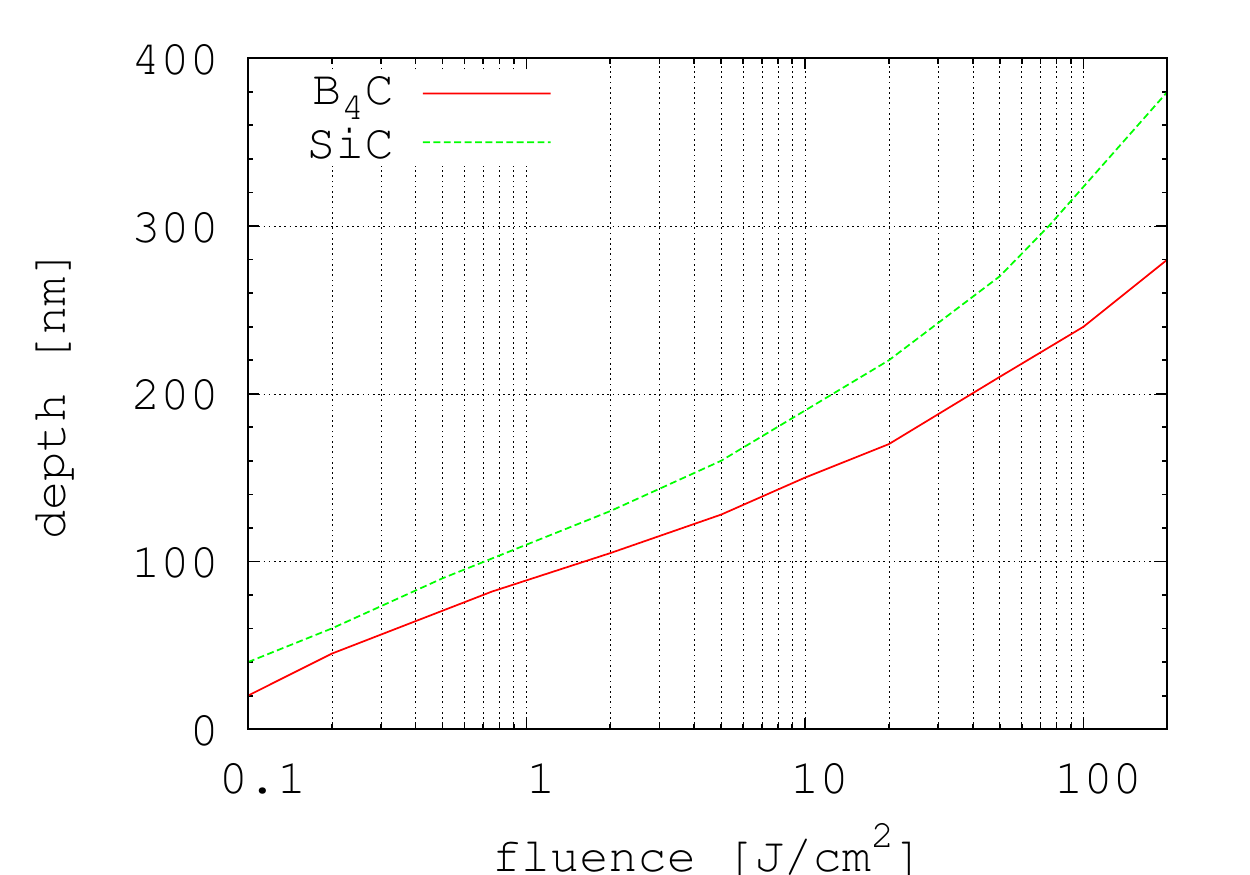}
\caption{Estimated crater depth for $\mathrm{B_{4}C}$ (red) and SiC (blue) based on the heating profile after the pulse, using the defined threshold for non-thermal ablation. The laser wavelength is 32.5 nm and the pulse length is 25 fs. The decrease in photoionization is almost compensated for by the inverse bremsstrahlung, resulting in an exponential dependence between depth and fluence up to about 10 $\mathrm{J/cm^2}$. For higher fluences the crater depths increase faster due to the decreased opacity.}\label{fig:depth_32nm}
\end{figure}

Figure \ref{fig:depth_32nm} shows the estimated crater depth for $\mathrm{B_{4}C}$ (red) and SiC (blue) as functions of fluence based on the simulated heating profiles, using the threshold for non-thermal ablation. The region below 3 $\mathrm{J/cm^2}$ is the same as the blue curves in figure \ref{fig:depth}(a) and (b). The effect of the reduced opacity has a marginal effect on the estimated crater depth for fluences below 10 $\mathrm{J/cm^2}$, as confirmed by the exponential dependence on fluence. Assuming a constant opacity $\kappa$, we expect an attenuation of the incident photon flux $I_0$: $I=I_0 e^{-\kappa \rho d}$, where $d$ the depth into the material and $\rho$ is the density of the material. Such a dependence would result in a straight line for the log-plot of figure \ref{fig:depth_32nm}. We conclude that the rate of inverse bremsstrahlung absorption is almost high enough to compensate for the decrease in the photoionization cross section. However, a moderate decrease in the total opacity can be identified above 10 $\mathrm{J/cm^2}$, especially for SiC (blue).

\subsection{Predictions at 6 nm wavelength}

In this section we compare the results of simulations in the vacuum-ultraviolet region with those of simulations in the mid-soft X-ray region. We choose a wavelength close to 6 nm since this will be available for experiments in the next stage of the FLASH soft X-ray free-electron laser \cite{Ackermann2007}. This study does not consider resonance heating where the laser wavelength overlaps with the absorption-peaks resulting from the bound-bound contributions. We hence choose to simulate the interaction of 5.9 nm radiation with $\mathrm{B_{4}C}$.

\begin{figure}[htb]
\includegraphics[width=7cm]{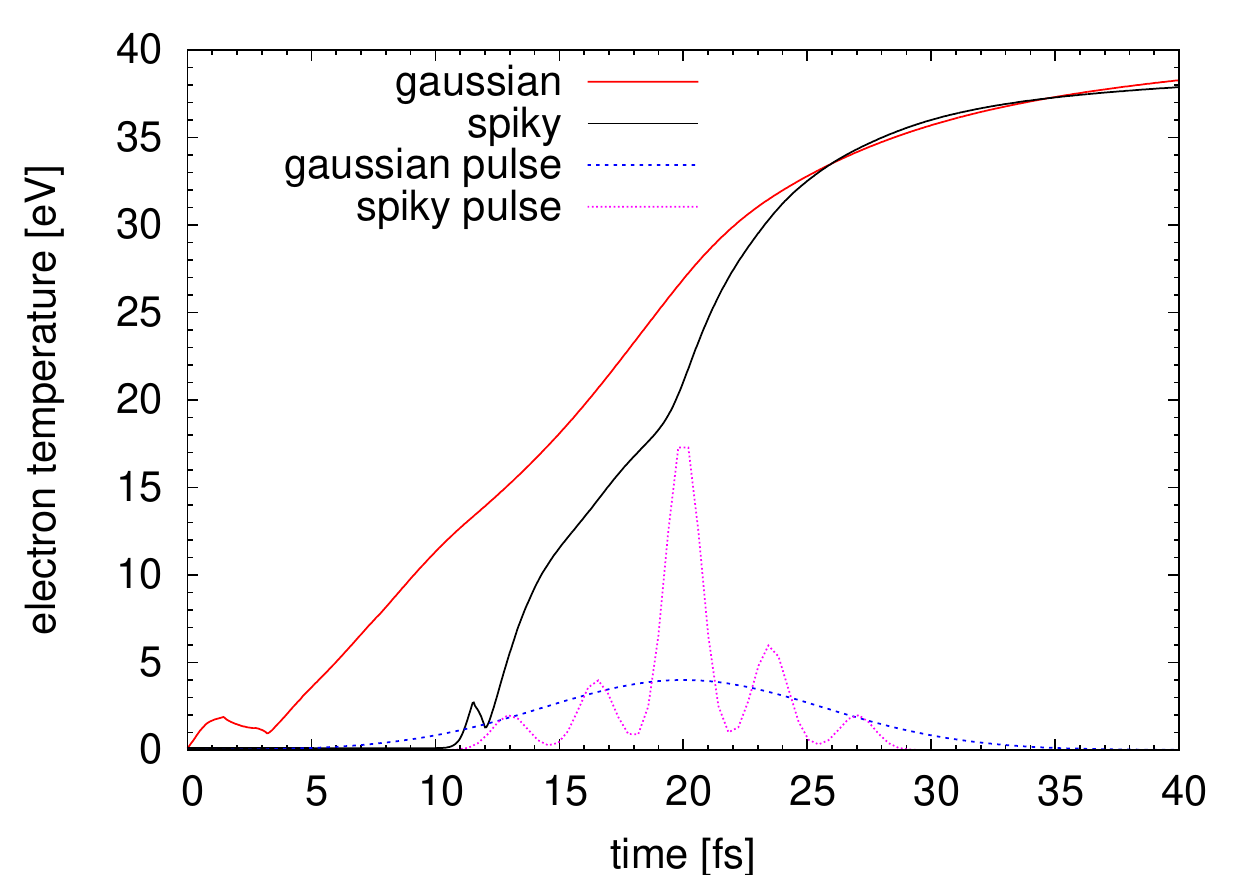}
\caption{Temporal evolution of the electron temperature at the surface with a Gaussian pulse profile (red) and a spiky pulse profile (black), at a wavelength of 5.9 nm, a fluence of 200$\mathrm{J/cm^2}$  and a pulse length of 15 fs.  The corresponding pulse shapes are shown in blue (Gaussian) and magenta (spiky), with an arbitrary intensity. Both pulses contain the same number of photons, and result in the same final electron temperature.}\label{fig:spikes}
\end{figure}

The temporal profile of the FEL pulse is expected to contain more pronounced spikes for shorter wavelength \cite{Salin2006}. To check how such a spiky profile affects the ionization dynamics, we simulate the heating of $\mathrm{B_{4}C}$ using both a Gaussian pulse and a pulse with a spiky temporal profile made to resemble the predictions by \citet{Salin2006}. The two pulses have equal energy (200 $\mathrm{J/cm^2}$) and pulse length (15 fs), and their shapes are shown in figure \ref{fig:spikes}. The Gaussian pulse (blue) induces a time-dependent electron temperature described by the red curve, while the spiky pulse (magenta) corresponds to the black temperature-curve. The temperatures follow different paths during the first part of the simulation, but converge to the same temperature during the second part of the simulation. Hence figure \ref{fig:spikes} supports the use of Gaussian pulse as a valid approximation in the following analysis. 

\begin{figure}[htb]
\includegraphics[width=7cm]{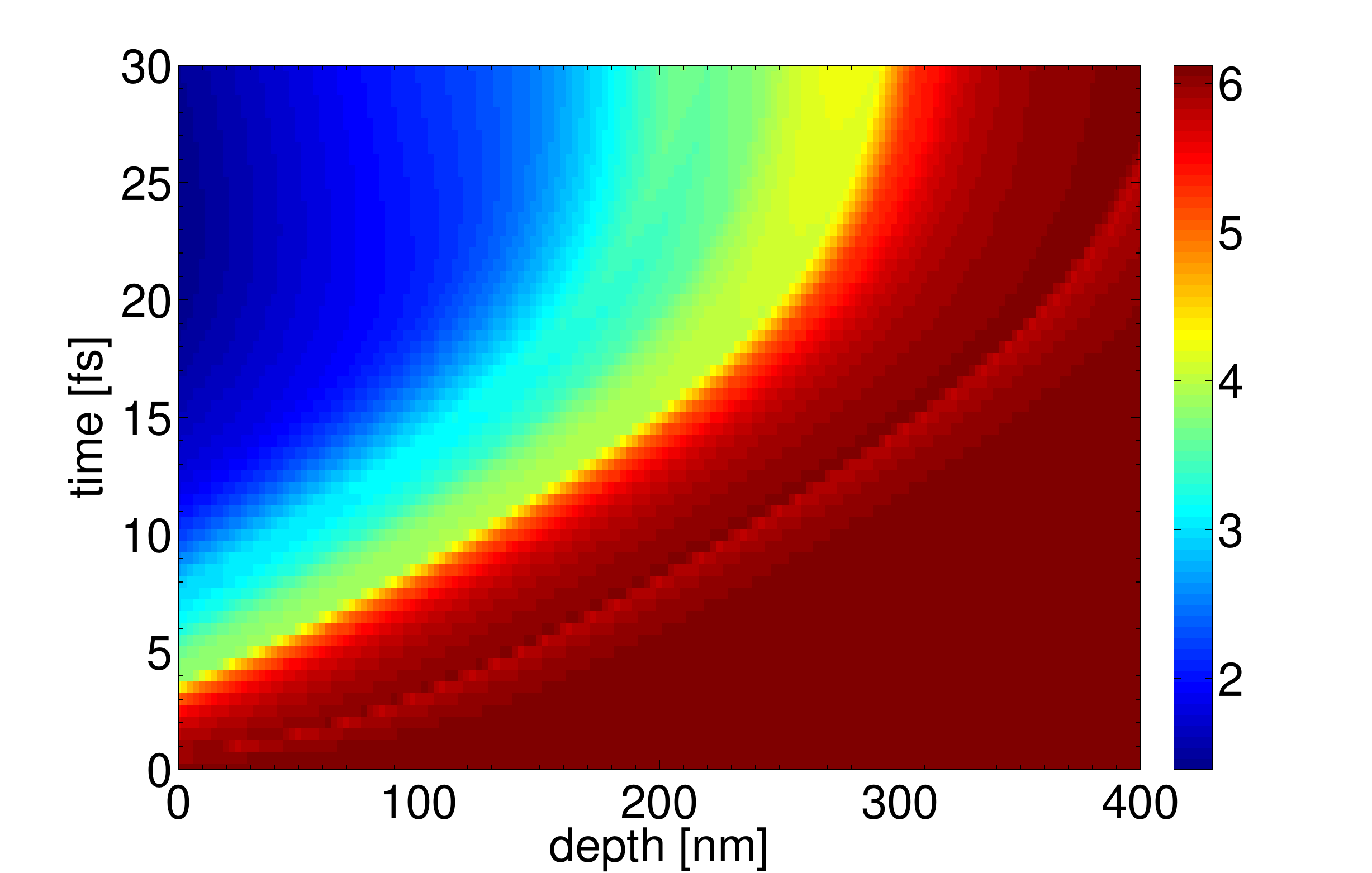}
\caption{Temporal evolution of the opacity as function of material depth for $\mathrm{B_{4}C}$. The pulse has a wavelength of 5.9 nm, a fluence of 200 $\mathrm{J/cm^2}$ and a pulse length of 15 fs. The opacity decreases by almost a factor of four due to the weak inverse bremsstrahlung.}\label{fig:b4c_op_6nm}
\end{figure}

Figure \ref{fig:b4c_op_6nm} shows the temporal evolution of the opacity as a function of material depth for a 5.9 nm, 200 $\mathrm{J/cm^2}$ Gaussian pulse. At this wavelength the photons are energetic enough to ionize the boron K-shell directly, but not the carbon K-shell.  The incident photons interact mainly with the core electrons in boron, with photoionizations followed by fast impact ionization of the valence electrons of both boron and carbon.  The free electrons are heated to a temperature of about 40 eV, which is too low for significant impact excitation or ionization of the K-shell electrons. In the case of 32.5 nm irradiation at the same fluence (described above), the photons interact with the valence electrons. The free electrons are heated to a temperature of about 100 eV.  This temperature is high enough for the electrons in the high-energy tail of the electron energy distribution to cause some excitation and ionization of the K-shell electrons in both boron and carbon.  The cold opacity at 5.9 nm is very similar to that at 32.5 nm, but during the pulse the opacity decreases by almost a factor of four, mainly due to the weaker inverse bremsstrahlung absorption at this relatively short wavelength.  

\begin{figure}[htb]
\includegraphics[width=7cm]{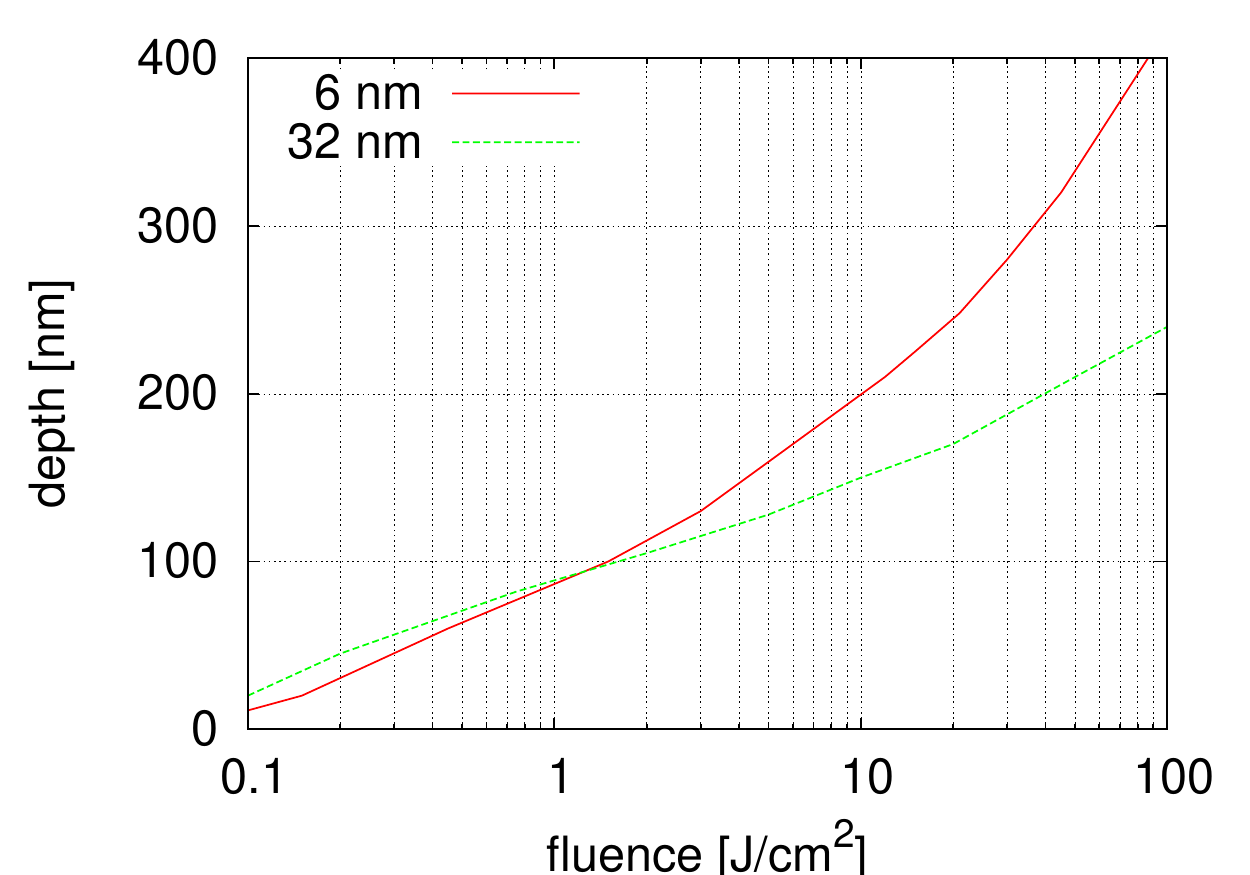}
\caption{Estimated crater depth for $\mathrm{B_{4}C}$ based on the heating profile after the pulse, using the defined threshold for non-thermal ablation. The 32.5 nm case is shown for comparison. There is a drastic decrease in opacity due to depopulation of the boron K-shell and the weak inverse bremsstrahlung, leading to an increase in the predicted rate of ablated material.}\label{fig:b4c_6nm}
\end{figure}

The consequence of the decrease of the opacity at 5.9 nm on the estimated crater depth is shown in figure \ref{fig:b4c_6nm}. For fluences up to about 2 $\mathrm{J/cm^2}$ the estimated heating depths are very similar to the 32.5 nm case, but for higher fluences the rate of ablated material is predicted to increase. 

We note that in the resonant region of the absorption spectrum, there will be line contributions from bound-bound transitions in the plasma, further complicating the absorption mechanisms. We expect considerable broadening of the lines at solid density, especially for low Z materials like $\mathrm{B_{4}C}$. In the case of Si, we expect the lines to result in wavelength regions of significantly increased absorption. Such an increase could affect the mechanisms of ablation and possibly increase the threshold fluence for melt expulsion, allowing more control over the crater depth and providing a cleaner ablation at higher fluences (due to the suppression of melt expulsion).

\section{Conclusions}

The advent of intense femtosecond pulses in the X-ray regime opens up a new branch of laser physics, where the interaction with materials proceeds by different physical processes compared to the optical regime. Beside the inherent interest of this regime, a quantitative understanding of the interaction is critical to find the optimal working parameters for demonstrated potential applications \cite{chapman2006a,Chalupsky2007}. We have addressed this issue by performing simulations of the heating using a plasma code that follows the atomic populations in time throughout the material.  

In conclusion, we have simulated the interaction of intense ultrafast X-ray pulses with the materials boron carbide ($\mathrm{B_{4}C}$), silicon carbide (SiC) and silicon using a non-local thermodynamic equilibrium (NLTE) radiation transfer code. We chose pulse parameters to match recent experiments at FLASH \cite{Ackermann2007}, and produced predictions for the next stage of operation of that facility \cite{Salin2006}. Our simulations show that at 32.5 nm wavelength, the decrease in opacity due to photoionization is partly compensated for by increased inverse bremsstrahlung absorption, leading to a near-exponential energy deposition in the material. At a wavelength of 6 nm this effect weakens, resulting in deeper energy deposition for fluences above 2 $\mathrm{J/cm^2}$. Crater depths can be estimated from the simulated heating profiles and the thresholds for ultrafast melting. To evaluate the model and investigate the mechanisms of ablation, we include a comparison to experimental crater depth data \cite{Hau-Riege2007d} for fluences up to 2.2 $\mathrm{J/cm^2}$. The experimental data for Si, and the 1D heating simulations support the idea proposed by \citet{Stojanovic2006} that melt expulsion leads to an increase in the crater depth at fluences above $\approx$ 1 $\mathrm{J/cm^2}$. Our simulations also indicate that, for these pulse parameters, an NLTE treatment is required for fluences above approximately 80 $\mathrm{J/cm^2}$.

\begin{acknowledgments}

This work was supported by the Swedish Research Council through the Center of Excellence in FEL-Studies at Uppsala University and the US Department of Energy through the Stanford Linear Accelerator Center.

\end{acknowledgments}

\newpage %Just because of unusual number of tables stacked at end
\bibliography{laserplasma}% Produces the bibliography via BibTeX.

\begin{thebibliography}{47}
\expandafter\ifx\csname natexlab\endcsname\relax\def\natexlab#1{#1}\fi
\expandafter\ifx\csname bibnamefont\endcsname\relax
  \def\bibnamefont#1{#1}\fi
\expandafter\ifx\csname bibfnamefont\endcsname\relax
  \def\bibfnamefont#1{#1}\fi
\expandafter\ifx\csname citenamefont\endcsname\relax
  \def\citenamefont#1{#1}\fi
\expandafter\ifx\csname url\endcsname\relax
  \def\url#1{\texttt{#1}}\fi
\expandafter\ifx\csname urlprefix\endcsname\relax\def\urlprefix{URL }\fi
\providecommand{\bibinfo}[2]{#2}
\providecommand{\eprint}[2][]{\url{#2}}

\bibitem[{\citenamefont{Ackermann et~al.}(2007)\citenamefont{Ackermann, Asova,
  Ayvazyan, Azima, Baboi, B\"ahr, Balandin, Beutner, Brandt, Bolzmann
  et~al.}}]{Ackermann2007}
\bibinfo{author}{\bibfnamefont{W.}~\bibnamefont{Ackermann}},
  \bibinfo{author}{\bibfnamefont{G.}~\bibnamefont{Asova}},
  \bibinfo{author}{\bibfnamefont{V.}~\bibnamefont{Ayvazyan}},
  \bibinfo{author}{\bibfnamefont{A.}~\bibnamefont{Azima}},
  \bibinfo{author}{\bibfnamefont{N.}~\bibnamefont{Baboi}},
  \bibinfo{author}{\bibfnamefont{J.}~\bibnamefont{B\"ahr}},
  \bibinfo{author}{\bibfnamefont{V.}~\bibnamefont{Balandin}},
  \bibinfo{author}{\bibfnamefont{B.}~\bibnamefont{Beutner}},
  \bibinfo{author}{\bibfnamefont{A.}~\bibnamefont{Brandt}},
  \bibinfo{author}{\bibfnamefont{A.}~\bibnamefont{Bolzmann}},
  \bibnamefont{et~al.}, \bibinfo{journal}{Nature Photonics}
  \textbf{\bibinfo{volume}{1}}, \bibinfo{pages}{336} (\bibinfo{year}{2007}).

\bibitem[{\citenamefont{Ayvazyan et~al.}(2006)\citenamefont{Ayvazyan, Baboi,
  B\"ahr, Balandin, Beutner, Brandt, Bohnet, Bolzmann, Brinkmann, Brovko
  et~al.}}]{Ayvazyan2006}
\bibinfo{author}{\bibfnamefont{V.}~\bibnamefont{Ayvazyan}},
  \bibinfo{author}{\bibfnamefont{N.}~\bibnamefont{Baboi}},
  \bibinfo{author}{\bibfnamefont{J.}~\bibnamefont{B\"ahr}},
  \bibinfo{author}{\bibfnamefont{V.}~\bibnamefont{Balandin}},
  \bibinfo{author}{\bibfnamefont{B.}~\bibnamefont{Beutner}},
  \bibinfo{author}{\bibfnamefont{A.}~\bibnamefont{Brandt}},
  \bibinfo{author}{\bibfnamefont{I.}~\bibnamefont{Bohnet}},
  \bibinfo{author}{\bibfnamefont{A.}~\bibnamefont{Bolzmann}},
  \bibinfo{author}{\bibfnamefont{R.}~\bibnamefont{Brinkmann}},
  \bibinfo{author}{\bibfnamefont{O.}~\bibnamefont{Brovko}},
  \bibnamefont{et~al.}, \bibinfo{journal}{Eur. Phys. J. D}
  \textbf{\bibinfo{volume}{37}}, \bibinfo{pages}{297} (\bibinfo{year}{2006}).

\bibitem[{\citenamefont{{LCLS Design Study-Report}}(1998)}]{LCLS}
\bibinfo{author}{\bibnamefont{{LCLS Design Study-Report}}},
  \bibinfo{journal}{SLAC-R-593, UC-414}  (\bibinfo{year}{1998}).

\bibitem[{\citenamefont{McPherson et~al.}(1987)\citenamefont{McPherson, Gibson,
  Jara, Johan, Luk, McInstyre, Boyer, and Rhodes}}]{McPherson1987}
\bibinfo{author}{\bibfnamefont{A.}~\bibnamefont{McPherson}},
  \bibinfo{author}{\bibfnamefont{G.}~\bibnamefont{Gibson}},
  \bibinfo{author}{\bibfnamefont{H.}~\bibnamefont{Jara}},
  \bibinfo{author}{\bibfnamefont{U.}~\bibnamefont{Johan}},
  \bibinfo{author}{\bibfnamefont{T.~S.} \bibnamefont{Luk}},
  \bibinfo{author}{\bibfnamefont{I.~A.} \bibnamefont{McInstyre}},
  \bibinfo{author}{\bibfnamefont{K.}~\bibnamefont{Boyer}}, \bibnamefont{and}
  \bibinfo{author}{\bibfnamefont{C.~K.} \bibnamefont{Rhodes}},
  \bibinfo{journal}{J. Opt. Soc. Am. B} \textbf{\bibinfo{volume}{4}},
  \bibinfo{pages}{595} (\bibinfo{year}{1987}).

\bibitem[{\citenamefont{Ferray et~al.}(1988)\citenamefont{Ferray, L'Hullier,
  Li, and L.~A.~Lompre}}]{Ferray1988}
\bibinfo{author}{\bibfnamefont{M.}~\bibnamefont{Ferray}},
  \bibinfo{author}{\bibfnamefont{A.}~\bibnamefont{L'Hullier}},
  \bibinfo{author}{\bibfnamefont{X.~F.} \bibnamefont{Li}}, \bibnamefont{and}
  \bibinfo{author}{\bibfnamefont{G.~M.} \bibnamefont{L.~A.~Lompre}},
  \bibinfo{journal}{J. Phys. B: At. Mol. Opt.Phys.}
  \textbf{\bibinfo{volume}{21}}, \bibinfo{pages}{31} (\bibinfo{year}{1988}).

\bibitem[{\citenamefont{Takahashi et~al.}(2000)\citenamefont{Takahashi,
  Nabekawa, and Midorikawa}}]{Takahashi2000}
\bibinfo{author}{\bibfnamefont{E.}~\bibnamefont{Takahashi}},
  \bibinfo{author}{\bibfnamefont{N.}~\bibnamefont{Nabekawa}}, \bibnamefont{and}
  \bibinfo{author}{\bibfnamefont{K.}~\bibnamefont{Midorikawa}},
  \bibinfo{journal}{Opt. Lett.} \textbf{\bibinfo{volume}{27}},
  \bibinfo{pages}{1920} (\bibinfo{year}{2000}).

\bibitem[{\citenamefont{Geddes et~al.}(2004)\citenamefont{Geddes, Cs.~Toth,
  Esarey, Schroeder, Bruhwiler, Nieter, Cary, and Leemans}}]{Leemans2004}
\bibinfo{author}{\bibfnamefont{C.~G.~R.} \bibnamefont{Geddes}},
  \bibinfo{author}{\bibfnamefont{J.~v.~T.} \bibnamefont{Cs.~Toth}},
  \bibinfo{author}{\bibfnamefont{E.}~\bibnamefont{Esarey}},
  \bibinfo{author}{\bibfnamefont{C.~B.} \bibnamefont{Schroeder}},
  \bibinfo{author}{\bibfnamefont{D.}~\bibnamefont{Bruhwiler}},
  \bibinfo{author}{\bibfnamefont{C.}~\bibnamefont{Nieter}},
  \bibinfo{author}{\bibfnamefont{J.}~\bibnamefont{Cary}}, \bibnamefont{and}
  \bibinfo{author}{\bibfnamefont{W.~P.} \bibnamefont{Leemans}},
  \bibinfo{journal}{Nature} \textbf{\bibinfo{volume}{431}},
  \bibinfo{pages}{538} (\bibinfo{year}{2004}).

\bibitem[{\citenamefont{Faure et~al.}(2004)\citenamefont{Faure, Glinec, Puhkov,
  Kiselev, Gordienko, Lefebvre, Rousseau, Burgy, and Malka}}]{Faure2004}
\bibinfo{author}{\bibfnamefont{J.}~\bibnamefont{Faure}},
  \bibinfo{author}{\bibfnamefont{Y.}~\bibnamefont{Glinec}},
  \bibinfo{author}{\bibfnamefont{A.}~\bibnamefont{Puhkov}},
  \bibinfo{author}{\bibfnamefont{S.}~\bibnamefont{Kiselev}},
  \bibinfo{author}{\bibfnamefont{S.}~\bibnamefont{Gordienko}},
  \bibinfo{author}{\bibfnamefont{E.}~\bibnamefont{Lefebvre}},
  \bibinfo{author}{\bibfnamefont{J.~P.} \bibnamefont{Rousseau}},
  \bibinfo{author}{\bibfnamefont{F.}~\bibnamefont{Burgy}}, \bibnamefont{and}
  \bibinfo{author}{\bibfnamefont{V.}~\bibnamefont{Malka}},
  \bibinfo{journal}{Nature} \textbf{\bibinfo{volume}{431}},
  \bibinfo{pages}{541} (\bibinfo{year}{2004}).

\bibitem[{\citenamefont{Gr\"uner et~al.}(2007)\citenamefont{Gr\"uner, Becker,
  Schramm, Eichner, Fuchs, Weingartner, Habs, ter vehn, Geissler, Ferrario
  et~al.}}]{Gruner2007}
\bibinfo{author}{\bibfnamefont{F.}~\bibnamefont{Gr\"uner}},
  \bibinfo{author}{\bibfnamefont{S.}~\bibnamefont{Becker}},
  \bibinfo{author}{\bibfnamefont{U.}~\bibnamefont{Schramm}},
  \bibinfo{author}{\bibfnamefont{T.}~\bibnamefont{Eichner}},
  \bibinfo{author}{\bibfnamefont{M.}~\bibnamefont{Fuchs}},
  \bibinfo{author}{\bibfnamefont{R.}~\bibnamefont{Weingartner}},
  \bibinfo{author}{\bibfnamefont{D.}~\bibnamefont{Habs}},
  \bibinfo{author}{\bibfnamefont{J.~M.} \bibnamefont{ter vehn}},
  \bibinfo{author}{\bibfnamefont{M.}~\bibnamefont{Geissler}},
  \bibinfo{author}{\bibfnamefont{M.}~\bibnamefont{Ferrario}},
  \bibnamefont{et~al.}, \bibinfo{journal}{Appl. Phys. B.}
  \textbf{\bibinfo{volume}{86}}, \bibinfo{pages}{431} (\bibinfo{year}{2007}).

\bibitem[{\citenamefont{Neutze et~al.}(2000)\citenamefont{Neutze, Wouts,
  van~der Spoel, Weckert, and Hajdu}}]{Neutze2000a}
\bibinfo{author}{\bibfnamefont{R.}~\bibnamefont{Neutze}},
  \bibinfo{author}{\bibfnamefont{R.}~\bibnamefont{Wouts}},
  \bibinfo{author}{\bibfnamefont{D.}~\bibnamefont{van~der Spoel}},
  \bibinfo{author}{\bibfnamefont{E.}~\bibnamefont{Weckert}}, \bibnamefont{and}
  \bibinfo{author}{\bibfnamefont{J.}~\bibnamefont{Hajdu}},
  \bibinfo{journal}{Nature} \textbf{\bibinfo{volume}{406}},
  \bibinfo{pages}{752} (\bibinfo{year}{2000}).

\bibitem[{\citenamefont{Bergh et~al.}(2007)\citenamefont{Bergh, Huldt,
  Timneanu, and Hajdu}}]{Bergh2007}
\bibinfo{author}{\bibfnamefont{M.}~\bibnamefont{Bergh}},
  \bibinfo{author}{\bibfnamefont{G.}~\bibnamefont{Huldt}},
  \bibinfo{author}{\bibfnamefont{N.}~\bibnamefont{Timneanu}}, \bibnamefont{and}
  \bibinfo{author}{\bibfnamefont{J.}~\bibnamefont{Hajdu}}, \bibinfo{journal}{J.
  Struct. Biol.}  (\bibinfo{year}{2007}).

\bibitem[{\citenamefont{Sokolowski-Tinten
  et~al.}(1995)\citenamefont{Sokolowski-Tinten, Bialkowski, and von~der
  Linde}}]{Tinten1995}
\bibinfo{author}{\bibfnamefont{K.}~\bibnamefont{Sokolowski-Tinten}},
  \bibinfo{author}{\bibfnamefont{J.}~\bibnamefont{Bialkowski}},
  \bibnamefont{and} \bibinfo{author}{\bibfnamefont{D.}~\bibnamefont{von~der
  Linde}}, \bibinfo{journal}{Phys. Rev. {\bf B}} \textbf{\bibinfo{volume}{51}},
  \bibinfo{pages}{14186} (\bibinfo{year}{1995}).

\bibitem[{\citenamefont{Lindenberg et~al.}(2005)\citenamefont{Lindenberg,
  Larsson, Sokolowski-Tinten, Gaffney, Blome, Synnergren, Sheppard, Caleman,
  MacPhee, Weinstein et~al.}}]{Lindenberg2005a}
\bibinfo{author}{\bibfnamefont{A.~M.} \bibnamefont{Lindenberg}},
  \bibinfo{author}{\bibfnamefont{J.}~\bibnamefont{Larsson}},
  \bibinfo{author}{\bibfnamefont{K.}~\bibnamefont{Sokolowski-Tinten}},
  \bibinfo{author}{\bibfnamefont{K.~J.} \bibnamefont{Gaffney}},
  \bibinfo{author}{\bibfnamefont{C.}~\bibnamefont{Blome}},
  \bibinfo{author}{\bibfnamefont{O.}~\bibnamefont{Synnergren}},
  \bibinfo{author}{\bibfnamefont{J.}~\bibnamefont{Sheppard}},
  \bibinfo{author}{\bibfnamefont{C.}~\bibnamefont{Caleman}},
  \bibinfo{author}{\bibfnamefont{A.~G.} \bibnamefont{MacPhee}},
  \bibinfo{author}{\bibfnamefont{D.}~\bibnamefont{Weinstein}},
  \bibnamefont{et~al.}, \bibinfo{journal}{Science}
  \textbf{\bibinfo{volume}{308}}, \bibinfo{pages}{392} (\bibinfo{year}{2005}).

\bibitem[{\citenamefont{Town et~al.}(1994)\citenamefont{Town, Bell, and
  Rose}}]{Town1994}
\bibinfo{author}{\bibfnamefont{R.~P.~J.} \bibnamefont{Town}},
  \bibinfo{author}{\bibfnamefont{A.~R.} \bibnamefont{Bell}}, \bibnamefont{and}
  \bibinfo{author}{\bibfnamefont{S.~J.} \bibnamefont{Rose}},
  \bibinfo{journal}{Phys. Rev. {\bf E}} \textbf{\bibinfo{volume}{50}},
  \bibinfo{pages}{1431} (\bibinfo{year}{1994}).

\bibitem[{\citenamefont{Jiang and Tsai}(2004)}]{Jiang2004}
\bibinfo{author}{\bibfnamefont{L.}~\bibnamefont{Jiang}} \bibnamefont{and}
  \bibinfo{author}{\bibfnamefont{H.~L.} \bibnamefont{Tsai}},
  \bibinfo{journal}{J. Phys. D: Appl. Phys.} \textbf{\bibinfo{volume}{37}},
  \bibinfo{pages}{1492} (\bibinfo{year}{2004}).

\bibitem[{\citenamefont{Lorazo et~al.}(2006)\citenamefont{Lorazo, Lewis, and
  Meunier}}]{lorazo2006}
\bibinfo{author}{\bibfnamefont{P.}~\bibnamefont{Lorazo}},
  \bibinfo{author}{\bibfnamefont{L.~J.} \bibnamefont{Lewis}}, \bibnamefont{and}
  \bibinfo{author}{\bibfnamefont{M.}~\bibnamefont{Meunier}},
  \bibinfo{journal}{Phys. Rev. {\bf B}} \textbf{\bibinfo{volume}{73}},
  \bibinfo{pages}{134108} (\bibinfo{year}{2006}).

\bibitem[{\citenamefont{Fajardo et~al.}(2004)\citenamefont{Fajardo, Zeitoun,
  and Gauthier}}]{Fajardo2004}
\bibinfo{author}{\bibfnamefont{M.}~\bibnamefont{Fajardo}},
  \bibinfo{author}{\bibfnamefont{P.}~\bibnamefont{Zeitoun}}, \bibnamefont{and}
  \bibinfo{author}{\bibfnamefont{J.-C.} \bibnamefont{Gauthier}},
  \bibinfo{journal}{Eur. Phys. J. D} \textbf{\bibinfo{volume}{29}},
  \bibinfo{pages}{69} (\bibinfo{year}{2004}).

\bibitem[{\citenamefont{Theobald et~al.}(1999)\citenamefont{Theobald,
  H\"assner, Kingham, Sauerbrey, Fehr, Gericke, Schlenges, Kraeft, and
  Ishikawa}}]{Theobald1999}
\bibinfo{author}{\bibfnamefont{W.}~\bibnamefont{Theobald}},
  \bibinfo{author}{\bibfnamefont{R.}~\bibnamefont{H\"assner}},
  \bibinfo{author}{\bibfnamefont{R.}~\bibnamefont{Kingham}},
  \bibinfo{author}{\bibfnamefont{R.}~\bibnamefont{Sauerbrey}},
  \bibinfo{author}{\bibfnamefont{R.}~\bibnamefont{Fehr}},
  \bibinfo{author}{\bibfnamefont{D.~O.} \bibnamefont{Gericke}},
  \bibinfo{author}{\bibfnamefont{M.}~\bibnamefont{Schlenges}},
  \bibinfo{author}{\bibfnamefont{W.~D.} \bibnamefont{Kraeft}},
  \bibnamefont{and} \bibinfo{author}{\bibfnamefont{K.}~\bibnamefont{Ishikawa}},
  \bibinfo{journal}{Phys. Rev. {\bf E}} \textbf{\bibinfo{volume}{59}},
  \bibinfo{pages}{3544} (\bibinfo{year}{1999}).

\bibitem[{\citenamefont{Chapman et~al.}(2007)\citenamefont{Chapman, Hau-Riege,
  Bogan, Bajt, Barty, Boutet, Marchesini, Frank, Woods, Benner
  et~al.}}]{Chapman2007b}
\bibinfo{author}{\bibfnamefont{H.}~\bibnamefont{Chapman}},
  \bibinfo{author}{\bibfnamefont{S.}~\bibnamefont{Hau-Riege}},
  \bibinfo{author}{\bibfnamefont{M.}~\bibnamefont{Bogan}},
  \bibinfo{author}{\bibfnamefont{S.}~\bibnamefont{Bajt}},
  \bibinfo{author}{\bibfnamefont{A.}~\bibnamefont{Barty}},
  \bibinfo{author}{\bibfnamefont{S.}~\bibnamefont{Boutet}},
  \bibinfo{author}{\bibfnamefont{S.}~\bibnamefont{Marchesini}},
  \bibinfo{author}{\bibfnamefont{M.}~\bibnamefont{Frank}},
  \bibinfo{author}{\bibfnamefont{B.}~\bibnamefont{Woods}},
  \bibinfo{author}{\bibfnamefont{W.}~\bibnamefont{Benner}},
  \bibnamefont{et~al.}, \bibinfo{journal}{Nature}
  \textbf{\bibinfo{volume}{448}}, \bibinfo{pages}{676} (\bibinfo{year}{2007}).

\bibitem[{\citenamefont{Hau-Riege
  et~al.}(2007{\natexlab{a}})\citenamefont{Hau-Riege, London, Chapman, and
  Bergh}}]{hau-riege2007c}
\bibinfo{author}{\bibfnamefont{S.~P.} \bibnamefont{Hau-Riege}},
  \bibinfo{author}{\bibfnamefont{R.~A.} \bibnamefont{London}},
  \bibinfo{author}{\bibfnamefont{H.~N.} \bibnamefont{Chapman}},
  \bibnamefont{and} \bibinfo{author}{\bibfnamefont{M.}~\bibnamefont{Bergh}},
  \bibinfo{journal}{Phys. Rev. {\bf E}} p. \bibinfo{pages}{in press}
  (\bibinfo{year}{2007}{\natexlab{a}}).

\bibitem[{\citenamefont{Hau-Riege
  et~al.}(2007{\natexlab{b}})\citenamefont{Hau-Riege, London, Bionta, McKerman,
  Baker, Krzywinski, Sobierajski, Nietubyc, Pelka, Jurek
  et~al.}}]{Hau-Riege2007d}
\bibinfo{author}{\bibfnamefont{S.~P.} \bibnamefont{Hau-Riege}},
  \bibinfo{author}{\bibfnamefont{R.~A.} \bibnamefont{London}},
  \bibinfo{author}{\bibfnamefont{R.~M.} \bibnamefont{Bionta}},
  \bibinfo{author}{\bibfnamefont{M.~A.} \bibnamefont{McKerman}},
  \bibinfo{author}{\bibfnamefont{S.~L.} \bibnamefont{Baker}},
  \bibinfo{author}{\bibfnamefont{J.}~\bibnamefont{Krzywinski}},
  \bibinfo{author}{\bibfnamefont{R.}~\bibnamefont{Sobierajski}},
  \bibinfo{author}{\bibfnamefont{R.}~\bibnamefont{Nietubyc}},
  \bibinfo{author}{\bibfnamefont{J.}~\bibnamefont{Pelka}},
  \bibinfo{author}{\bibfnamefont{M.}~\bibnamefont{Jurek}},
  \bibnamefont{et~al.}, \bibinfo{journal}{Appl. Phys. Lett.}
  \textbf{\bibinfo{volume}{90}}, \bibinfo{pages}{173128}
  (\bibinfo{year}{2007}{\natexlab{b}}).

\bibitem[{\citenamefont{Theobald et~al.}(1996)\citenamefont{Theobald,
  H\"assner, W\"ulker, and Sauerbrey}}]{Theobald1996}
\bibinfo{author}{\bibfnamefont{W.}~\bibnamefont{Theobald}},
  \bibinfo{author}{\bibfnamefont{R.}~\bibnamefont{H\"assner}},
  \bibinfo{author}{\bibfnamefont{C.}~\bibnamefont{W\"ulker}}, \bibnamefont{and}
  \bibinfo{author}{\bibfnamefont{R.}~\bibnamefont{Sauerbrey}},
  \bibinfo{journal}{Phys. Rev. Lett.} \textbf{\bibinfo{volume}{77}},
  \bibinfo{pages}{298} (\bibinfo{year}{1996}).

\bibitem[{\citenamefont{Dobosz et~al.}(2005)\citenamefont{Dobosz, Doumy,
  Stabile, D'oliveira, Monot, Réau, H{\"u}ller, and Martin}}]{Dobosz2005}
\bibinfo{author}{\bibfnamefont{S.}~\bibnamefont{Dobosz}},
  \bibinfo{author}{\bibfnamefont{G.}~\bibnamefont{Doumy}},
  \bibinfo{author}{\bibfnamefont{H.}~\bibnamefont{Stabile}},
  \bibinfo{author}{\bibfnamefont{P.}~\bibnamefont{D'oliveira}},
  \bibinfo{author}{\bibfnamefont{P.}~\bibnamefont{Monot}},
  \bibinfo{author}{\bibfnamefont{F.}~\bibnamefont{Réau}},
  \bibinfo{author}{\bibfnamefont{S.}~\bibnamefont{H{\"u}ller}},
  \bibnamefont{and} \bibinfo{author}{\bibfnamefont{P.}~\bibnamefont{Martin}},
  \bibinfo{journal}{Phys. Rev. Lett.} \textbf{\bibinfo{volume}{95}},
  \bibinfo{pages}{25001} (\bibinfo{year}{2005}).

\bibitem[{\citenamefont{Chapman et~al.}(2006)\citenamefont{Chapman, Barty,
  Bogan, Boutet, Frank, Hau-Riege, Marchesini, Woods, Bajt, London
  et~al.}}]{chapman2006a}
\bibinfo{author}{\bibfnamefont{H.~N.} \bibnamefont{Chapman}},
  \bibinfo{author}{\bibfnamefont{A.}~\bibnamefont{Barty}},
  \bibinfo{author}{\bibfnamefont{M.~J.} \bibnamefont{Bogan}},
  \bibinfo{author}{\bibfnamefont{S.}~\bibnamefont{Boutet}},
  \bibinfo{author}{\bibfnamefont{M.}~\bibnamefont{Frank}},
  \bibinfo{author}{\bibfnamefont{S.~P.} \bibnamefont{Hau-Riege}},
  \bibinfo{author}{\bibfnamefont{S.}~\bibnamefont{Marchesini}},
  \bibinfo{author}{\bibfnamefont{B.~W.} \bibnamefont{Woods}},
  \bibinfo{author}{\bibfnamefont{S.}~\bibnamefont{Bajt}},
  \bibinfo{author}{\bibfnamefont{R.~A.} \bibnamefont{London}},
  \bibnamefont{et~al.}, \bibinfo{journal}{Nat. Phys.}
  \textbf{\bibinfo{volume}{12}}, \bibinfo{pages}{839} (\bibinfo{year}{2006}).

\bibitem[{\citenamefont{Chalupsky et~al.}(2007)\citenamefont{Chalupsky, Juha,
  Kuba, Cihelka, Hajkova, Koptyaev, Krasa, Velyhan, Bergh, Caleman
  et~al.}}]{Chalupsky2007}
\bibinfo{author}{\bibfnamefont{J.}~\bibnamefont{Chalupsky}},
  \bibinfo{author}{\bibfnamefont{L.}~\bibnamefont{Juha}},
  \bibinfo{author}{\bibfnamefont{J.}~\bibnamefont{Kuba}},
  \bibinfo{author}{\bibfnamefont{J.}~\bibnamefont{Cihelka}},
  \bibinfo{author}{\bibfnamefont{V.}~\bibnamefont{Hajkova}},
  \bibinfo{author}{\bibfnamefont{S.}~\bibnamefont{Koptyaev}},
  \bibinfo{author}{\bibfnamefont{J.}~\bibnamefont{Krasa}},
  \bibinfo{author}{\bibfnamefont{A.}~\bibnamefont{Velyhan}},
  \bibinfo{author}{\bibfnamefont{M.}~\bibnamefont{Bergh}},
  \bibinfo{author}{\bibfnamefont{C.}~\bibnamefont{Caleman}},
  \bibnamefont{et~al.}, \bibinfo{journal}{Optics Express}
  \textbf{\bibinfo{volume}{15}}, \bibinfo{pages}{6036} (\bibinfo{year}{2007}).

\bibitem[{\citenamefont{Stampfli and Benneman}(1994)}]{Stampfli1994}
\bibinfo{author}{\bibfnamefont{P.}~\bibnamefont{Stampfli}} \bibnamefont{and}
  \bibinfo{author}{\bibfnamefont{K.~H.} \bibnamefont{Benneman}},
  \bibinfo{journal}{Phys. Rev. {\bf B}} \textbf{\bibinfo{volume}{49}},
  \bibinfo{pages}{7299} (\bibinfo{year}{1994}).

\bibitem[{\citenamefont{Eidmann et~al.}(2000)\citenamefont{Eidmann, ter Vehn,
  and Schlegel}}]{Eidmann2000}
\bibinfo{author}{\bibfnamefont{K.}~\bibnamefont{Eidmann}},
  \bibinfo{author}{\bibfnamefont{J.~M.} \bibnamefont{ter Vehn}},
  \bibnamefont{and} \bibinfo{author}{\bibfnamefont{T.}~\bibnamefont{Schlegel}},
  \bibinfo{journal}{Phys. Rev. {\bf E}} \textbf{\bibinfo{volume}{62}},
  \bibinfo{pages}{1202} (\bibinfo{year}{2000}).

\bibitem[{\citenamefont{Scott and Mayle}(1994)}]{scott1992}
\bibinfo{author}{\bibfnamefont{H.~A.} \bibnamefont{Scott}} \bibnamefont{and}
  \bibinfo{author}{\bibfnamefont{R.~W.} \bibnamefont{Mayle}},
  \bibinfo{journal}{Appl. Phys. B.} \textbf{\bibinfo{volume}{58}},
  \bibinfo{pages}{35} (\bibinfo{year}{1994}).

\bibitem[{\citenamefont{Scott}(2001)}]{scott2001}
\bibinfo{author}{\bibfnamefont{H.~A.} \bibnamefont{Scott}},
  \bibinfo{journal}{JQSRT} \textbf{\bibinfo{volume}{71}}, \bibinfo{pages}{689}
  (\bibinfo{year}{2001}).

\bibitem[{\citenamefont{More}(1982)}]{More1982}
\bibinfo{author}{\bibfnamefont{R.}~\bibnamefont{More}},
  \bibinfo{journal}{JQSRT} \textbf{\bibinfo{volume}{27}}, \bibinfo{pages}{345}
  (\bibinfo{year}{1982}).

\bibitem[{\citenamefont{Henke et~al.}(1993)\citenamefont{Henke, Gullikson, and
  Davis}}]{Henke1993}
\bibinfo{author}{\bibfnamefont{B.}~\bibnamefont{Henke}},
  \bibinfo{author}{\bibfnamefont{E.}~\bibnamefont{Gullikson}},
  \bibnamefont{and} \bibinfo{author}{\bibfnamefont{J.}~\bibnamefont{Davis}},
  \bibinfo{journal}{Atom. Nucl. Data Tabl.} \textbf{\bibinfo{volume}{54}},
  \bibinfo{pages}{181} (\bibinfo{year}{1993}).

\bibitem[{\citenamefont{Stewart and Pyatt}(1966)}]{stewart1966}
\bibinfo{author}{\bibfnamefont{J.~C.} \bibnamefont{Stewart}} \bibnamefont{and}
  \bibinfo{author}{\bibfnamefont{K.~D.} \bibnamefont{Pyatt}},
  \bibinfo{journal}{Apj.} \textbf{\bibinfo{volume}{144}}, \bibinfo{pages}{1203}
  (\bibinfo{year}{1966}).

\bibitem[{\citenamefont{Nantel et~al.}(1998)\citenamefont{Nantel, Ma, Gu, Coté,
  Itatani, and Umstadter}}]{nantel1998}
\bibinfo{author}{\bibfnamefont{M.}~\bibnamefont{Nantel}},
  \bibinfo{author}{\bibfnamefont{G.}~\bibnamefont{Ma}},
  \bibinfo{author}{\bibfnamefont{S.}~\bibnamefont{Gu}},
  \bibinfo{author}{\bibfnamefont{C.~Y.} \bibnamefont{Coté}},
  \bibinfo{author}{\bibfnamefont{J.}~\bibnamefont{Itatani}}, \bibnamefont{and}
  \bibinfo{author}{\bibfnamefont{D.}~\bibnamefont{Umstadter}},
  \bibinfo{journal}{Phys. Rev. Lett.} \textbf{\bibinfo{volume}{20}},
  \bibinfo{pages}{4442} (\bibinfo{year}{1998}).

\bibitem[{\citenamefont{Dawson and Oberman}(1962)}]{Dawson1962}
\bibinfo{author}{\bibfnamefont{J.}~\bibnamefont{Dawson}} \bibnamefont{and}
  \bibinfo{author}{\bibfnamefont{C.}~\bibnamefont{Oberman}},
  \bibinfo{journal}{Phys. Fluids} \textbf{\bibinfo{volume}{5}},
  \bibinfo{pages}{517} (\bibinfo{year}{1962}).

\bibitem[{\citenamefont{Gericke et~al.}(2002)\citenamefont{Gericke, Murillo,
  and Schlanges}}]{Schlanges2002}
\bibinfo{author}{\bibfnamefont{D.}~\bibnamefont{Gericke}},
  \bibinfo{author}{\bibfnamefont{M.~S.} \bibnamefont{Murillo}},
  \bibnamefont{and}
  \bibinfo{author}{\bibfnamefont{M.}~\bibnamefont{Schlanges}},
  \bibinfo{journal}{Phys. Rev. {\bf E}} \textbf{\bibinfo{volume}{65}},
  \bibinfo{pages}{036418} (\bibinfo{year}{2002}).

\bibitem[{\citenamefont{Spitzer}(1956)}]{Spitzer1956}
\bibinfo{author}{\bibfnamefont{L.}~\bibnamefont{Spitzer}},
  \bibinfo{journal}{Physics of Fully Ionized Gases (Interscience Publishers,
  New York}  (\bibinfo{year}{1956}).

\bibitem[{\citenamefont{Brabec and Krausz}(2000)}]{Brabec2000}
\bibinfo{author}{\bibfnamefont{T.}~\bibnamefont{Brabec}} \bibnamefont{and}
  \bibinfo{author}{\bibfnamefont{F.}~\bibnamefont{Krausz}},
  \bibinfo{journal}{Reviews of Modern Physics} \textbf{\bibinfo{volume}{72}},
  \bibinfo{pages}{545} (\bibinfo{year}{2000}).

\bibitem[{\citenamefont{Delone and Krainov}(2000)}]{Delone2000}
\bibinfo{author}{\bibfnamefont{N.~B.} \bibnamefont{Delone}} \bibnamefont{and}
  \bibinfo{author}{\bibfnamefont{V.~P.} \bibnamefont{Krainov}},
  p.~\bibinfo{pages}{92} (\bibinfo{year}{2000}).

\bibitem[{\citenamefont{Saldin et~al.}(2006)\citenamefont{Saldin,
  Schneidmiller, and Yurkov}}]{Salin2006}
\bibinfo{author}{\bibfnamefont{E.}~\bibnamefont{Saldin}},
  \bibinfo{author}{\bibfnamefont{E.}~\bibnamefont{Schneidmiller}},
  \bibnamefont{and} \bibinfo{author}{\bibfnamefont{M.}~\bibnamefont{Yurkov}},
  \bibinfo{journal}{Nucl. Inst. Meth. Phys. Res. A}
  \textbf{\bibinfo{volume}{562}}, \bibinfo{pages}{472} (\bibinfo{year}{2006}).

\bibitem[{\citenamefont{Ziaja et~al.}(2005)\citenamefont{Ziaja, London, and
  Hajdu}}]{Ziaja2005a}
\bibinfo{author}{\bibfnamefont{B.}~\bibnamefont{Ziaja}},
  \bibinfo{author}{\bibfnamefont{R.~A.} \bibnamefont{London}},
  \bibnamefont{and} \bibinfo{author}{\bibfnamefont{J.}~\bibnamefont{Hajdu}},
  \bibinfo{journal}{J. Appl. Phys.} \textbf{\bibinfo{volume}{97}},
  \bibinfo{pages}{064905} (\bibinfo{year}{2005}).

\bibitem[{\citenamefont{Timneanu et~al.}(2004)\citenamefont{Timneanu, Caleman,
  Hajdu, and van~der Spoel}}]{Timneanu2004a}
\bibinfo{author}{\bibfnamefont{N.}~\bibnamefont{Timneanu}},
  \bibinfo{author}{\bibfnamefont{C.}~\bibnamefont{Caleman}},
  \bibinfo{author}{\bibfnamefont{J.}~\bibnamefont{Hajdu}}, \bibnamefont{and}
  \bibinfo{author}{\bibfnamefont{D.}~\bibnamefont{van~der Spoel}},
  \bibinfo{journal}{Chem. Phys.} \textbf{\bibinfo{volume}{299}},
  \bibinfo{pages}{277} (\bibinfo{year}{2004}).

\bibitem[{\citenamefont{Rousse et~al.}(2001)\citenamefont{Rousse, Rischel,
  Fourmaux, Uschermann, Sebban, Grillon, Balcou, F\"orster, Geindre, Audebert
  et~al.}}]{Rousse2001}
\bibinfo{author}{\bibfnamefont{A.}~\bibnamefont{Rousse}},
  \bibinfo{author}{\bibfnamefont{C.}~\bibnamefont{Rischel}},
  \bibinfo{author}{\bibfnamefont{S.}~\bibnamefont{Fourmaux}},
  \bibinfo{author}{\bibfnamefont{I.}~\bibnamefont{Uschermann}},
  \bibinfo{author}{\bibfnamefont{S.}~\bibnamefont{Sebban}},
  \bibinfo{author}{\bibfnamefont{G.}~\bibnamefont{Grillon}},
  \bibinfo{author}{\bibfnamefont{P.}~\bibnamefont{Balcou}},
  \bibinfo{author}{\bibfnamefont{E.}~\bibnamefont{F\"orster}},
  \bibinfo{author}{\bibfnamefont{J.~P.} \bibnamefont{Geindre}},
  \bibinfo{author}{\bibfnamefont{P.}~\bibnamefont{Audebert}},
  \bibnamefont{et~al.}, \bibinfo{journal}{Nature}
  \textbf{\bibinfo{volume}{410}}, \bibinfo{pages}{65} (\bibinfo{year}{2001}).

\bibitem[{\citenamefont{More et~al.}(1988)\citenamefont{More, Warren, Young,
  and Zimmerman}}]{More1988}
\bibinfo{author}{\bibfnamefont{R.~M.} \bibnamefont{More}},
  \bibinfo{author}{\bibfnamefont{K.~H.} \bibnamefont{Warren}},
  \bibinfo{author}{\bibfnamefont{D.~A.} \bibnamefont{Young}}, \bibnamefont{and}
  \bibinfo{author}{\bibfnamefont{G.~B.} \bibnamefont{Zimmerman}},
  \bibinfo{journal}{Phys. Fluids} \textbf{\bibinfo{volume}{31}},
  \bibinfo{pages}{3059} (\bibinfo{year}{1988}).

\bibitem[{\citenamefont{Martynyuk}(1974)}]{Martynyuk1974}
\bibinfo{author}{\bibfnamefont{M.~M.} \bibnamefont{Martynyuk}},
  \bibinfo{journal}{Sov. Phys. Tech. Phys.} \textbf{\bibinfo{volume}{19}},
  \bibinfo{pages}{793} (\bibinfo{year}{1974}).

\bibitem[{\citenamefont{Stojanovic et~al.}(2006)\citenamefont{Stojanovic,
  von~der Linde, Sokolowski-Tinten, Zastrau, Perner, F\"orster, Sobierajski,
  Jurek, Klinger, Pelka et~al.}}]{Stojanovic2006}
\bibinfo{author}{\bibfnamefont{N.}~\bibnamefont{Stojanovic}},
  \bibinfo{author}{\bibfnamefont{D.}~\bibnamefont{von~der Linde}},
  \bibinfo{author}{\bibfnamefont{K.}~\bibnamefont{Sokolowski-Tinten}},
  \bibinfo{author}{\bibfnamefont{U.}~\bibnamefont{Zastrau}},
  \bibinfo{author}{\bibfnamefont{F.}~\bibnamefont{Perner}},
  \bibinfo{author}{\bibfnamefont{E.}~\bibnamefont{F\"orster}},
  \bibinfo{author}{\bibfnamefont{R.}~\bibnamefont{Sobierajski}},
  \bibinfo{author}{\bibfnamefont{R.~N.~M.} \bibnamefont{Jurek}},
  \bibinfo{author}{\bibfnamefont{D.}~\bibnamefont{Klinger}},
  \bibinfo{author}{\bibfnamefont{J.}~\bibnamefont{Pelka}},
  \bibnamefont{et~al.}, \bibinfo{journal}{Appl. Phys. Lett.}
  \textbf{\bibinfo{volume}{89}}, \bibinfo{pages}{241909}
  (\bibinfo{year}{2006}).

\bibitem[{\citenamefont{Mima}(1994)}]{Mima1994}
\bibinfo{author}{\bibfnamefont{K.}~\bibnamefont{Mima}},
  \emph{\bibinfo{title}{Laser Plasma Theory and Simulation}}
  (\bibinfo{publisher}{harwood academic publishers}, \bibinfo{year}{1994}),
  \bibinfo{edition}{1st} ed., ISBN \bibinfo{isbn}{3-7186-5489-X}.

\bibitem[{\citenamefont{Butler et~al.}(2004)\citenamefont{Butler, Gonsalves,
  McKenna, Spence, Hooker, Sebban, Mocek, Bettaibi, and Cros}}]{Butler2004}
\bibinfo{author}{\bibfnamefont{A.}~\bibnamefont{Butler}},
  \bibinfo{author}{\bibfnamefont{A.~J.} \bibnamefont{Gonsalves}},
  \bibinfo{author}{\bibfnamefont{C.~M.} \bibnamefont{McKenna}},
  \bibinfo{author}{\bibfnamefont{D.~J.} \bibnamefont{Spence}},
  \bibinfo{author}{\bibfnamefont{S.~M.} \bibnamefont{Hooker}},
  \bibinfo{author}{\bibfnamefont{S.}~\bibnamefont{Sebban}},
  \bibinfo{author}{\bibfnamefont{T.}~\bibnamefont{Mocek}},
  \bibinfo{author}{\bibfnamefont{I.}~\bibnamefont{Bettaibi}}, \bibnamefont{and}
  \bibinfo{author}{\bibfnamefont{B.}~\bibnamefont{Cros}},
  \bibinfo{journal}{Phys. Rev. {\bf A}} \textbf{\bibinfo{volume}{70}},
  \bibinfo{pages}{023821} (\bibinfo{year}{2004}).

\end{thebibliography}

\end{document}